\documentclass[11pt]{article}

\usepackage[utf8]{inputenc}
\usepackage[T1]{fontenc}
\usepackage{lmodern}
\usepackage{amsmath,amssymb}
\usepackage{graphicx}
\usepackage[margin=1in]{geometry}
\usepackage{hyperref}
\usepackage{booktabs}
\usepackage{multirow}
\usepackage[table]{xcolor}
\definecolor{hm1}{HTML}{D73027}  % strong red (most negative)
\definecolor{hm2}{HTML}{F46D43}
\definecolor{hm3}{HTML}{FDAE61}
\definecolor{hm4}{HTML}{FEE08B}
\definecolor{hm5}{HTML}{FFFFBF}  % neutral center
\definecolor{hm6}{HTML}{D9EF8B}
\definecolor{hm7}{HTML}{A6D96A}
\definecolor{hm8}{HTML}{66BD63}
\definecolor{hm9}{HTML}{1A9850}  % strong green (most positive)
\usepackage{algorithm}
\usepackage{algpseudocode}
\usepackage{enumitem}
\usepackage{natbib}
\usepackage{url}
\usepackage{placeins}
\usepackage{pdflscape}
\usepackage{float}
\usepackage{listings}
\hypersetup{
  colorlinks=true,
  linkcolor=blue!60!black,
  citecolor=blue!60!black,
  urlcolor=blue!60!black,
  plainpages=false,
  pdfpagelabels,
  hypertexnames=false
}

\title{Epicure: Multidimensional Flavor Structure in\\Food Ingredient Embeddings\\[6pt]
\large LLM-Augmented Data Curation Reveals Culturally and Perceptually\\Grounded Dimensions in Food Embeddings}

\author{
  Jakub Radzikowski\\
  KAIKAKU.AI\\
  \texttt{jakub@kaikaku.ai}
  \and
  Josef Chen\\
  KAIKAKU.AI\\
  \texttt{josef@kaikaku.ai}
}

\date{}

\begin{document}
\maketitle

% ============================================================
\begin{abstract}
A chef's intuition about flavor, texture, and cultural identity represents tacit knowledge that is difficult to articulate yet central to culinary practice. We show that this knowledge is already encoded in FlavorGraph's 300-dimensional ingredient embeddings, trained on recipe co-occurrence and food chemistry, and that it can be systematically recovered. An LLM-augmented curation pipeline consolidates 6,653 raw FlavorGraph ingredients into 1,032 canonical entries, substantially strengthening the recoverable structure.

We identify at least \textbf{fifteen} independently classifiable dimensions spanning taste, texture, geography, food processing, and culture. Seven exhibit statistically significant ordinal gradients ($\rho = 0.37$--$0.76$, all $p < .001$), from Scoville heat and climate latitude to the five basic tastes and NOVA processing level. Seven further dimensions, including sour, bitter, hardness, moisture, and fattiness, show significant binary separation (Cohen's $d = 0.79$--$1.26$, all $p < .001$). Cultural clustering across seven macro-regional cuisines achieves $k$NN purity of $0.43$ ($6.2\times$ lift over baseline), roughly doubling after curation.

Crucially, these findings are not artifacts of LLM labeling. Cross-validation against USDA FoodData Central laboratory measurements, involving no LLM at any stage, confirms five macronutrient dimensions ($\rho = 0.44$--$0.47$, all $p < .001$) and validates that taste axes track real chemical variation (sugars--sweet $\rho = 0.41$, sodium--salty $\rho = 0.29$). A hub ablation tracing signal provenance through FlavorGraph's 1,561 training compounds reveals that sweetness and saltiness arise entirely from recipe co-occurrence (sucrose and sodium are absent from the training data), while umami, protein, and fat have partial chemical pathways. All dimensions survive ten-fold cross-validation. These results establish that recipe co-occurrence data encodes structured, multidimensional culinary knowledge, and that data curation is the primary bottleneck for extracting it.

\end{abstract}

% ============================================================
\section{Introduction}
\label{sec:intro}

The global food system, from raw ingredients through formulated products to food service, is valued at roughly 10\% of world GDP~\cite{vannieuwkoop2019foodsystem}, with food imports alone projected to exceed USD~2 trillion annually~\cite{fao2024foodoutlook}, yet its core creative processes, menu development, recipe innovation, and flavor pairing, remain overwhelmingly manual, experience-driven, and difficult to scale. A chef's intuition about which ingredients complement each other in practical applications represents decades of accumulated knowledge that can be hard to externalise. The industry is ripe for a computational approach that augments, rather than replaces, human culinary creativity.

The computational study of ingredient compatibility began with the \emph{food pairing hypothesis}, the idea that ingredients sharing flavor compounds tend to pair well together. \cite{ahn2011flavor} formalized this by constructing a bipartite network linking ingredients to flavor compounds, then projecting it into a \emph{flavor network} where ingredients are connected by shared compounds. Their analysis of 56,498 recipes from three databases revealed a striking cultural divide: North American and Western European cuisines showed statistically significant over-representation of shared-compound pairings, while East Asian and Southern European cuisines showed the opposite pattern. This established that food pairing has structure amenable to computational analysis, though the shared-compound signal explains only part of what makes ingredients complement each other. IBM's Chef Watson~\cite{varshney2019chefwatson} further demonstrated that combining culinary science, food chemistry, and hedonic psychophysics with machine learning could achieve computational creativity, generating novel recipes by reasoning over ingredient compatibility data. Chef Watson showed that data-driven ingredient combination is viable at scale, but operated on hand-curated databases rather than learned embeddings.

\cite{park2021flavorgraph}, building on other work, addressed this gap with \emph{FlavorGraph}, a graph embedding method (metapath2vec~\cite{dong2017metapath2vec} with an additional chemical structure prediction layer) trained on a heterogeneous graph of approximately 8,300 total nodes (6,653 ingredient nodes and 1,561 flavor-compound nodes) and 147,000 edges derived from over one million recipes and 1,561 flavor molecules from FlavorDB~\cite{garg2018flavordb}. The resulting 300-dimensional embeddings capture both chemical affinities and recipe co-occurrence statistics, enabling food pairing recommendation that outperforms purely chemistry-based approaches. Other efforts in this direction include FoodKG~\cite{haussmann2019foodkg}, which integrates Recipe1M~\cite{marin2021recipe1m} data with USDA nutritional information and the FoodOn ontology~\cite{dooley2018foodon} into a 67-million-triple RDF knowledge graph supporting SPARQL-based recipe recommendation, and the broader field of computational gastronomy surveyed by \cite{bagler2024computational}, framing cooking as a creative process amenable to data-driven analysis across taste, nutrition, health, and sustainability dimensions.

However, a major gap remains between these research artifacts and tools that the hospitality industry can actually use. FlavorGraph's 6,653 ingredient nodes carry no categorization, no dietary metadata, and no mechanism for generating actionable outputs. Worse, the raw data is contaminated: of the 6,653 entries we select as our starting point, many are not food at all: kitchen paraphernalia (\emph{aluminum foil}, \emph{brown paper bag}, \emph{toothpick}), cleaning supplies (\emph{ammonia}, \emph{bleach}, \emph{borax}), brand-specific products (\emph{Betty Crocker Fudge Brownie Mix}, \emph{Campbell's Cream of Mushroom Soup}), and miscellaneous non-food items (\emph{bird seed}, \emph{beeswax}, \emph{plastic cup}); all with learned 300-dimensional embeddings. Among genuine ingredients, inconsistent naming is pervasive (``Kraft Sharp Cheddar Cheese'' and ``cheddar'' as separate entries), and 138 variants of ``beef'' (from ``80\% lean ground beef'' to ``beef tenderloin steak'') coexist as separate nodes, their embeddings reflecting different recipe co-occurrence contexts rather than coherent flavor profiles. These issues are typical of web-scraped recipe corpora~\cite{bien2020recipenlg} and the food domain more broadly. The ``data-centric AI'' paradigm~\cite{zha2025datacentric} argues that improving data quality often yields greater gains than improving model architecture, a perspective validated at scale by the FineWeb project~\cite{penedo2024fineweb}, which showed that classifier-based quality filtering of web text produces training corpora that outperform much larger uncurated datasets, and by DataComp~\cite{gadre2023datacomp}, which demonstrated that filtering CLIP's training data by quality metrics improved downstream performance more than scaling dataset size.

The dimensions that make a chef's intuition valuable: the balance of sweetness, the depth of umami, the calibration of heat, the degree to which ingredients have been processed, and how do they work together in a (inherently multimodal) food item are what \cite{polanyi1966tacit} termed \emph{tacit knowledge}: expertise that practitioners apply fluently but cannot fully articulate as explicit rules. \cite{collins2010tacit} further refined tacit knowledge into relational, somatic, and collective forms, distinguishing between knowledge that is not yet explicit and forms grounded in embodiment or social practice; \cite{dreyfus2001internet} emphasized the body's indispensable role in expert skill acquisition. A striking precedent for computational recovery of such implicit structure is the word2vec family~\cite{mikolov2013word2vec}: trained purely on word co-occurrence, these embeddings encode dimensions of meaning (gender, geography, and linguistic regularity) that were never labeled. The analogy to our work requires qualification: FlavorGraph embeddings were trained on both ingredient co-occurrence \emph{and} 1,561 flavor molecules from FlavorDB~\cite{garg2018flavordb}, so some chemical structure is expected. But if computational representations encode dimensions spanning nutritional composition, cultural identity, geographic origin, processing level, and texture, and if we can trace which signals arise from chemical compound edges vs.\ recipe co-occurrence, it would constitute evidence that recipe co-occurrence data encodes tacit culinary knowledge with measurable structure.

We present evidence that \emph{existing} FlavorGraph embeddings already encode precisely this kind of multidimensional structure: not only the chemical affinities present in the training data, but also nutritional composition, cultural clustering, the degree of ingredient processing, texture properties, and geographic climate origin. By tracing the chemical provenance of each signal through FlavorGraph's training compounds, we show that these dimensions arise from a gradient of sources: some primarily from recipe co-occurrence, others reinforced by chemical compound edges. Rather than building new embeddings or knowledge graphs, we strengthen these latent signals through Epicure, an LLM-augmented data curation pipeline that consolidates FlavorGraph's noisy ingredient space into 1,032 canonical entries. This substantially improves recoverable structure on the majority of dimensions tested, with the largest gains for Scoville, salty, latitude, and fattiness, indicating that current findings represent a lower bound on what the method can reveal. Our contributions are:

\begin{enumerate}[leftmargin=*]
\item \textbf{LLM-augmented data curation}: A multi-step pipeline that refines FlavorGraph's 6,653 selected ingredients into 1,032 canonical entries through progressive LLM-based normalization, semantic consolidation, human-in-the-loop quality control, embedding averaging, and categorization. Curation improves correlations on the majority of dimensions tested (up to $+0.24$ for Scoville), indicating that current results represent a lower bound on recoverable structure (Sections~\ref{sec:pipeline} and~\ref{sec:comparison}).

\item \textbf{Chemical and nutritional validation}: Independent cross-validation against USDA FoodData Central (712 matched ingredients) and FooDB compound measurements validates that embedding axes track real chemical variation (sugars--sweet $\rho=0.41$, glutamic acid--umami $\rho=0.24$) and encode five macronutrient dimensions (protein $\rho=0.47$, carbohydrates $\rho=0.47$, calories $\rho=0.47$, fiber $\rho=0.44$, fat $\rho=0.44$; all $p < .001$) defined entirely from laboratory measurements. A hub ablation with training compound provenance analysis traces each signal to its source: chemical edges, recipe co-occurrence, or both (Section~\ref{sec:chemical_validation}).

\item \textbf{Multidimensional embedding analysis}: Using LLM-driven food classification (Gemini~3.1 Pro with structured JSON output), we classify all ingredients along fifteen dimensions: five basic tastes, six texture properties, Scoville heat, climate zone, NOVA processing level, and cultural cuisine association. Seven dimensions show statistically significant ordinal correlations (all $p < .001$); seven further dimensions are verified through binary classification with significant embedding separation (Cohen's $d = 0.79$--$1.26$, all $p < .001$). The five taste axes are partially independent in the 300-dimensional space, with inter-axis angles revealing interpretable structure: umami--salty coupling, sweet--umami opposition, and sweet retaining the most unique signal after partial correlation analysis (Sections~\ref{sec:gt_results},~\ref{sec:texture_results}, and~\ref{sec:taste_geometry}).

\item \textbf{Cultural cluster analysis in native 300d space}: Using independently generated LLM annotations (7 distinctive macro-regional clusters), we show that $k$NN purity roughly doubles relative to a size-matched raw baseline after curation (Wilcoxon $p = 0.008$). UMAP analysis of the 300d embedding space enables visual interpretation of the cultural structure (Section~\ref{sec:cultural_results}).
\end{enumerate}

The UMAP visualization of the 300d ingredient embedding space, which benefits from interactive exploration, is available as a companion artifact available at \url{https://epicure-data.kaikaku.ai}

% ============================================================
\section{Methods}
\label{sec:methods}

% ============================================================
\subsection{Data Curation Pipeline}
\label{sec:pipeline}

Our pipeline transforms FlavorGraph's 6,653 selected ingredient entries, each associated with a 300-dimensional embedding vector, into 1,032 canonical entries with structured metadata.

\subsubsection{Three-Pass LLM Refinement (Steps 1--3)}

We employ three progressive passes of LLM-based name normalization and consolidation using Google Gemini 2.5 Flash\footnote{The data curation pipeline (Section~\ref{sec:pipeline}) was developed using Gemini 2.5 Flash. The deployed production system uses Gemini 3.1 Pro.} with structured JSON output.

\textbf{Step~1: Simplification.} Each ingredient name is cleaned by removing brand names, quantities, preparation details, and packaging information. We use Jaccard similarity-based semantic clustering to batch similar ingredients together before sending them to the LLM, improving consistency within clusters. For example, ``Kraft Sharp Cheddar Cheese (8 oz)'' becomes \texttt{cheddar\_cheese}, and ``Whole Grain Organic Brown Rice'' becomes \texttt{brown\_rice}.

\textbf{Step~2: Refinement.} Each simplified ingredient is classified by the LLM as \texttt{keep}, \texttt{remove} (non-food items, packaging materials, compound foods that are not ingredients)\footnote{An alternative approach for brand-specific products would be to decompose them into constituent base ingredients (e.g., resolving ``Campbell's Cream of Mushroom Soup'' to mushroom, cream, flour); we opted for removal as this decomposition introduces its own ambiguities and our focus was on canonical, single-ingredient entries.}, or \texttt{consolidate} (merge with a canonical entry).  This step also handles multi-language translation for entries in Spanish, French, Italian, German, and Portuguese using a dictionary of approximately 60 common translations, and applies consolidation patterns for cheese varieties, meat cuts, and sauce types.

\textbf{Step~3: Ultra-refinement.} A final aggressive consolidation pass applies approximately 450 lines of domain-specific rules encoded in the LLM prompt, covering: bread type unification (preserving sourdough and brioche as distinct), rice variety consolidation (preserving brown and wild rice, merging regular varieties), cultural spice preservation (retaining ras el hanout, za'atar, garam masala, and other culturally distinctive spice blends as distinct entries), and systematic consolidation of meat cuts, cheese varieties, pepper types, pasta shapes, and condiment variants.

Across three passes, the ingredient count is reduced from 6,653 to approximately 1,400 entries.

\subsubsection{Human-in-the-Loop Quality Control (Step~4)}

Automated LLM processing inevitably introduces errors: over-aggressive consolidation (merging ingredients that are culinarily distinct), under-consolidation (leaving obvious duplicates), and occasional hallucinated mappings. We address this with a custom editing interface that supports search, batch operations, and rapid review. A human curator reviews all entries, correcting LLM errors and making final consolidation decisions. This step typically adjusts 5--10\% of entries.

\subsubsection{Consolidation and Embedding Averaging (Steps~5--6)}

After quality control, we assign sequential canonical node IDs to the final set of unique ingredient names and generate a consolidation map tracking which original FlavorGraph node IDs merged into each canonical entry. For each consolidated ingredient, we compute a new 300-dimensional embedding as the arithmetic mean of all original FlavorGraph embeddings that mapped to it:
\begin{equation}
\mathbf{e}_{\text{consolidated}} = \frac{1}{|S|} \sum_{i \in S} \mathbf{e}_i
\label{eq:embedding_avg}
\end{equation}
where $S$ is the set of original node IDs that mapped to this consolidated ingredient. This averaging preserves the ``center of mass'' in flavor embedding space: if three cheese varieties are merged into a canonical \texttt{cheddar\_cheese}, the resulting embedding reflects the average flavor profile of those variants. We demonstrate in Section~\ref{sec:comparison} that the consolidated space yields stronger dimensional correlations than the raw space, primarily because consolidation removes duplicate entries that degrade aggregate statistics, and validate in Section~\ref{sec:cultural_results} that the resulting space preserves neighborhood relationships.

This averaging is theoretically grounded in results showing that weighted averages~\cite{arora2017embeddings} and simple means~\cite{wieting2016paraphrastic} of word embeddings can preserve compositional semantics. A more principled alternative: retraining FlavorGraph from scratch on the cleaned vocabulary, would avoid the information loss inherent in post-hoc averaging, but requires access to the original Recipe1M+~\cite{marin2021recipe1m} corpus with re-mapped annotations, placing it outside the scope of this work.

\subsubsection{Categorization and Dietary Classification (Step~7)}

Using Gemini 2.5 Flash with a 92-line categorization prompt, we assign each ingredient 1--3 categories from a taxonomy of 18 culinary categories: Meat, Fish, Seafood, Dairy, Veg, Fruit, Herbs, Spice, Nuts, Legumes, Grain, Fat, Sweet, Condiment, Beverage, Pantry, Protein, and Produce. This categorization system, utilized in our production app \url{http://epicure.kaikaku.ai}, follows \emph{culinary function} rather than biological taxonomy: tomatoes are categorized as Veg (used in savory cooking), rhubarb as Fruit (used in desserts), and corn as either Veg (fresh) or Grain (dried/processed). Each ingredient also receives binary vegetarian and vegan flags.

\subsubsection{Pairwise Similarity Computation (Step~8)}

We compute cosine similarity between all pairs of the 1,032 consolidated ingredient embeddings producing $\binom{1032}{2} = 531{,}996$ unique similarity scores.

\subsubsection{Pipeline Summary}

Table~\ref{tab:pipeline} summarizes the pipeline's quantitative characteristics.

\begin{table}[!htbp]
\centering
\caption{Data curation pipeline summary. The pipeline achieves 6.4$\times$ consolidation while preserving embedding semantics and adding structured metadata.}
\label{tab:pipeline}
\begin{tabular}{@{}lrrl@{}}
\toprule
\textbf{Stage} & \textbf{Input} & \textbf{Output} & \textbf{Key technique} \\
\midrule
Raw FlavorGraph & 6,653 & --- & 300D GNN embeddings \\
Step 1: Simplification & 6,653 & $\sim$4,800 & LLM + Jaccard clustering \\
Step 2: Refinement & $\sim$4,800 & $\sim$2,100 & LLM classification \\
Step 3: Ultra-refinement & $\sim$2,100 & $\sim$1,400 & Domain-specific rules \\
Step 4: Manual QC & $\sim$1,400 & $\sim$1,050 & Streamlit editor \\
Step 5: Consolidation & $\sim$1,050 & 1,032 & ID reassignment \\
Step 6: Embeddings & 1,032 & 1,032 & Mean averaging \\
Step 7: Categorization & 1,032 & 1,032 & LLM + 18 categories \\
Step 8: Similarity & 1,032 & 531,996 & Cosine similarity \\
\midrule
\textbf{Final output} & \multicolumn{3}{l}{1,032 ingredients $\times$ 300D embeddings + metadata} \\
\bottomrule
\end{tabular}
\end{table}

The final output consists of three artifacts: an ingredient list with categories and dietary flags, a $1{,}032 \times 300$ embedding matrix, and a table of 531,996 pairwise cosine similarities.

% ============================================================
\subsection{LLM-Driven Food Classification}
\label{sec:food_classification}

To test whether the 300-dimensional embeddings encode independently classifiable dimensions of culinary knowledge, we classify all ingredients along \textbf{15 dimensions} using Gemini~3.1 Pro\footnote{Model identifier: \texttt{gemini-3.1-pro-preview}. Crucially, both the curated set (1,032 ingredients) and the raw set (6,653 ingredients) are tagged \emph{independently}, enabling unbiased comparison.}

\textbf{Taste and chemistry dimensions (8).} (1)~\emph{Climate zone}: the climate in which the ingredient is primarily cultivated (Tropical, Subtropical, Mediterranean, Arid, Temperate, Continental, Subarctic), treated as a latitude-ordered ordinal variable (Tropical $\to$ Subarctic) in projection analyses. (2)~\emph{NOVA processing level}: integer 1--4 following the Monteiro classification~\cite{monteiro2019nova}. (3)~\emph{Umami level}: ordinal scale (none, low, moderate, high, very\_high). (4)~\emph{Scoville SHU}: estimated heat in Scoville Heat Units (0 for non-pungent ingredients). (5--8)~\emph{Sweet, salty, sour, bitter levels}: ordinal scale matching the umami classification.

\textbf{Texture dimensions (6).} Grounded in the ISO~11036 sensory texture profile standard~\cite{iso11036} and the Szczesniak classification~\cite{szczesniak1963texture}, each ingredient is classified \emph{as typically used in cooking}: (9)~\emph{Hardness} (liquid $\to$ very\_hard), (10)~\emph{Viscosity} (thin $\to$ very\_thick; N/A for solids), (11)~\emph{Crunchiness} (none $\to$ very\_high), (12)~\emph{Chewiness} (none $\to$ very\_high), (13)~\emph{Moisture} (dry $\to$ very\_wet), and (14)~\emph{Fattiness} (none $\to$ very\_high).

\textbf{Cultural dimension (1).} (15)~\emph{Cuisine association}: each ingredient is tagged with one of seven macro-regional cuisine clusters (Japanese, East Asian, Southeast Asian, South Asian, Latin American, Mediterranean, Northern/Atlantic) \emph{only} if it is a distinctive cultural marker; universal ingredients are left untagged. Details of the clustering protocol are in Section~\ref{sec:cultural_clusters}.

\textbf{Direct binary labels (7 dimensions).} For sour, bitter, and five texture dimensions (hardness, crunchiness, chewiness, moisture, fattiness), we run a second, independent yes/no classification pass (``Is this ingredient notably X? Yes/No'') using the same model and batching procedure. These direct binary labels are the primary inputs for binary separation analyses.

\textbf{Procedure.} Ingredients are batched (50 per request) and sent to the LLM with structured prompts (one ordinal multi-dimension prompt and one binary yes/no prompt for the 7 binary dimensions). The model returns structured JSON arrays conforming to predefined schemas, enforced via Gemini's \texttt{response\_mime\_type} parameter. A retry loop re-submits any untagged ingredients until 100\% coverage is achieved.

\textbf{Classification quality.} Spot-checking confirms expected classifications: cayenne (40,000~SHU) and habanero (250,000~SHU) receive high Scoville estimates while wasabi and horseradish correctly receive 0~SHU (their pungency is isothiocyanate-based, not capsaicin); miso, soy sauce, and parmesan are tagged very\_high umami; sugar and honey are very\_high sweet; lemon and vinegar are high sour; coffee and dark chocolate are high bitter.

\textbf{Methodological note.} These LLM-generated classifications serve as \emph{proxy labels} reflecting the model's training data, not as measured sensory or chemical ground truth. Any correlation between embedding projections and LLM labels demonstrates that two independently trained systems, FlavorGraph and Gemini, agree on the same dimensional structure, but does not constitute direct sensory validation. We provide independent chemical cross-validation for some of these using USDA and FooDB laboratory measurements in Section~\ref{sec:chemical_validation}.

% ============================================================
\subsection{Embedding Analysis Methods}
\label{sec:analysis_methods}

We use four statistical approaches to test whether the embeddings encode each classified dimension.

\textbf{Ordinal projection.} For dimensions with a natural ordering (tastes, NOVA, Scoville, latitude), we define an axis in the 300-dimensional space from the centroid of the lowest-level ingredients to the centroid of the highest-level ingredients. Each ingredient is projected onto this axis, and we compute Spearman rank correlation between the LLM-assigned rank and the projection value. This tests whether the embedding space encodes the ordinal structure of the dimension as a linear gradient. Because the axis is defined on the same data used to compute the correlation, reported $\rho$ values may be inflated. To quantify this, we report 10-fold cross-validated $\rho$ for all ordinal and binary dimensions (Tables~\ref{tab:gt_ordinal}, \ref{tab:extreme}, \ref{tab:texture_ordinal}, \ref{tab:nutritional}): the axis is defined from training-fold data only and correlation (or Cohen's $d$) is measured on the held-out fold, with 20 random repeats (200 total measurements per dimension).

\textbf{Per-ingredient categorical delta.} For categorical dimensions (climate zone), we compute each ingredient's mean cosine similarity to other members of its group (\emph{within}) minus its mean similarity to all other ingredients (\emph{cross}). The resulting per-ingredient deltas are independent observations suitable for a Wilcoxon signed-rank test. Cohen's $d$ on these deltas quantifies the effect size.

\textbf{Binary separation test.} For dimensions with direct binary labels (sour, bitter, hardness, crunchiness, chewiness, moisture, fattiness), we define an axis from the No-centroid to the Yes-centroid, project all ingredients, and report Mann-Whitney $U$ and Cohen's $d$.

\textbf{Axis geometry.} For both the five basic taste axes and the six texture axes, we compute inter-axis cosine similarity matrices. Ordinal axes use the low-pole to high-pole centroid direction; binary axes (sour, bitter, and five texture dimensions) use the No-centroid to Yes-centroid direction from direct LLM labels. Partial Spearman correlations (regressing out projections on the other axes via OLS residualization) quantify the degree of independence between dimensions. For cultural clusters, we additionally project the 7 cuisine centroids onto all classified dimension axes (taste, texture, Scoville, NOVA, latitude) and assess significance via permutation test ($N = 10{,}000$).

\textbf{Dimensionality reduction.} We project the 300-dimensional embeddings into three dimensions using UMAP~\cite{mcinnes2018umap} with cosine metric (\texttt{n\_neighbors}${}=32$, \texttt{min\_dist}${}=0.004$), selected via grid search over 16 configurations for tightest category separation. UMAP is used for visualization only; all quantitative claims are grounded in the native 300-dimensional space.

\textbf{$k$NN purity and cluster tightness.} For cultural clusters, we compute the fraction of each ingredient's $k=10$ nearest neighbors (cosine metric, 300d) that share the same cuisine tag, with random-baseline normalization and bootstrap subsampling (200 iterations) to control for pool size differences between raw and curated spaces. We additionally report intra-cluster pairwise cosine similarity and per-ingredient distance-to-centroid tests. The cuisine-tag protocol (distinctive-marker criterion, seven macro-regional clusters, and savoury-only restriction) is defined in Section~\ref{sec:cultural_clusters}, where it is presented alongside outcomes for interpretability.

% ============================================================
\subsection{Chemical and Nutritional Validation}
\label{sec:chemical_methods}

To test whether the embedding structure reflects genuine physicochemical properties rather than shared biases between FlavorGraph and the LLM classifier, we conduct two independent cross-validations using laboratory-measured data.

\textbf{Data sources.} We use USDA FoodData Central~\cite{usda2025fooddata} (Full Download, December 2025), filtering to Foundation Foods and SR Legacy foods for a total of 8,221 foods with measured nutrient values. For flavor compound concentrations, we use FooDB~\cite{foodb2020} (2020 CSV release), extracting L-glutamic acid and sugar compounds with sufficient coverage (sucrose, fructose, and glucose), with total sugars computed as their sum.

\textbf{Ingredient matching.} We match the 1,032 curated ingredients to USDA and FooDB entries using a three-layer pipeline (details in Appendix~\ref{sec:matching_pipeline}): (1)~rule-based matching via an inverted index over normalized name segments, with synonym mapping and preparation-state preference; (2)~embedding similarity using Gemini text embeddings (3,072-dim) with cosine-similarity ranking; and (3)~LLM validation of embedding candidates using Gemini~2.5~Flash with structured output. This yields 712 USDA matches (69.0\%) and 770 FooDB matches (74.6\%). Using an LLM for name matching does not compromise measurement independence: the LLM determines \emph{which} database row corresponds to a given ingredient name, while the nutrient or compound values are laboratory measurements.

\textbf{Cross-validation analysis.} For each USDA nutrient (sugars, sodium, water) or FooDB compound (total sugars: glucose + fructose + sucrose, glutamic acid), we project matched ingredients onto the corresponding LLM-defined taste axis and compute Spearman rank correlation between the projection and the measured value. For nutritional dimensions (protein, total fat, fiber, calories, carbohydrates), we define axes via tercile centroids of the measured values, entirely from USDA data, with no LLM involvement, and correlate the embedding projections with the measured nutrient content.

\textbf{Chemical hub ablation.} Of the 1,032 curated ingredients, 308 trace back to FlavorGraph's 416 ``chemical-hub'' nodes (those with direct compound edges in the training graph). The remaining 724 received chemical information only indirectly via metapath propagation. We report correlations separately for hub and non-hub subsets to assess whether each signal comes primarily from chemical training edges or from recipe co-occurrence.

% ============================================================
\section{Results}
\label{sec:results}

We apply the analysis methods of Section~\ref{sec:analysis_methods} to both the curated (1,032 ingredients) and raw (6,653 ingredients) embedding spaces, using the independent LLM classifications from Section~\ref{sec:food_classification}. All quantitative results are computed in the native 300-dimensional embedding space unless otherwise noted. Table~\ref{tab:cohorts} defines the analysis cohorts used throughout this section.

\begin{table}[!htbp]
\centering
\caption{Analysis cohorts. Each row defines a sample used in one or more analyses, with the inclusion rule and sections where it appears.}
\label{tab:cohorts}
\scriptsize
\begin{tabular}{@{}llrl@{}}
\toprule
\textbf{Cohort} & \textbf{Inclusion rule} & $n$ & \textbf{Used in} \\
\midrule
Curated (all)   & After curation pipeline & 1,032 & Secs.~\ref{sec:gt_results}--\ref{sec:texture_results} \\
Raw (all)       & All selected FlavorGraph nodes & 6,653 & LLM tagging (Sec.~\ref{sec:food_classification}) \\
Raw (categorized) & Raw nodes with back-projected categories & 4,804 & Raw-vs-curated comparison (Sec.~\ref{sec:comparison}) \\
Curated savoury & Curated, flavor profile = savoury & 589 & Culture analysis (Sec.~\ref{sec:cultural_results}) \\
Raw savoury     & Raw (categorized), flavor = savoury & 2,916 & Culture analysis (Sec.~\ref{sec:cultural_results}) \\
USDA matched    & Curated, matched to USDA & 712 & Chemical validation (Sec.~\ref{sec:chemical_validation}) \\
FooDB matched   & Curated, matched to FooDB & 770 & Chemical validation (Sec.~\ref{sec:chemical_validation}) \\
\bottomrule
\end{tabular}
\end{table}

We first evaluate which culinary dimensions are encoded in the native 300-dimensional space, then use UMAP geometry as qualitative context within the cultural analysis section (Section~\ref{sec:cultural_results}).

\FloatBarrier
% ============================================================
\subsection{Ordinal and Continuous Dimensions}
\label{sec:gt_results}

\begin{figure}[!htbp]
\centering
\includegraphics[width=\textwidth]{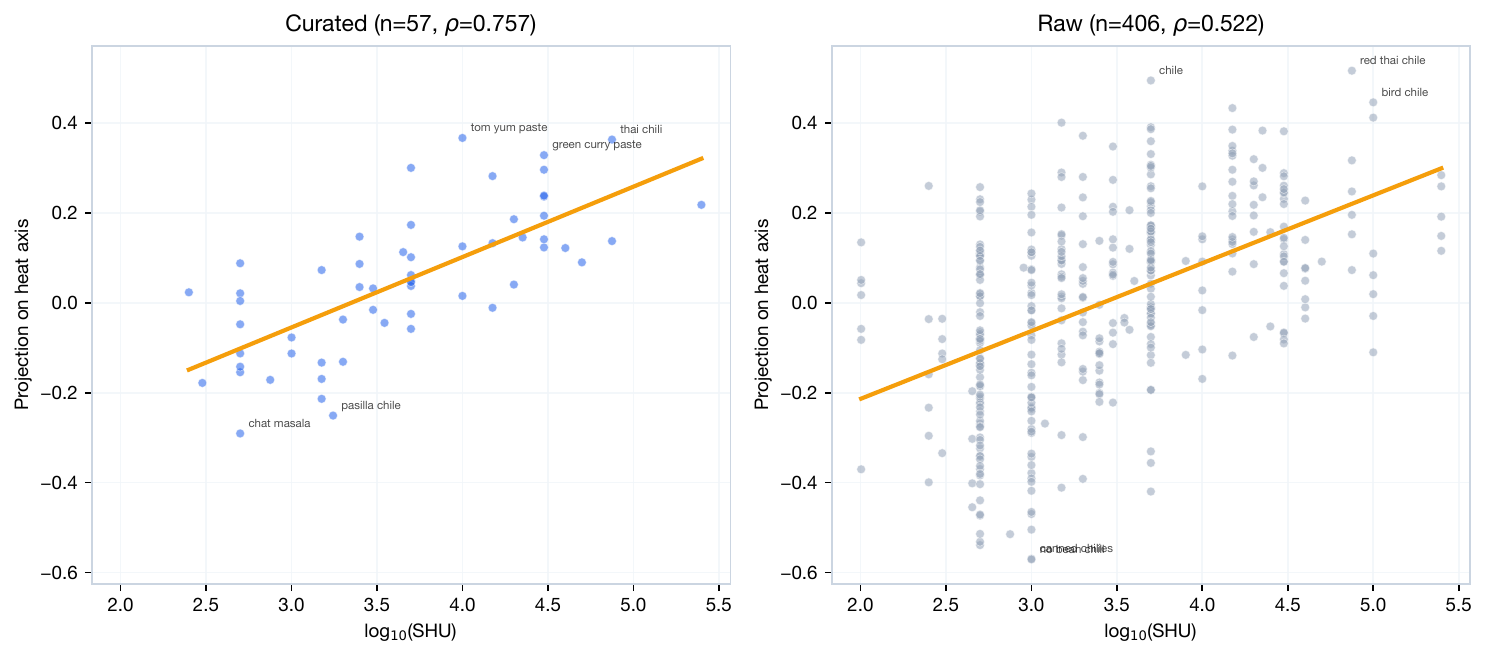}
\caption{Scoville heat scale: direct projection of spicy ingredients onto the heat axis (low-SHU to high-SHU tercile centroids) vs.\ $\log_{10}(\text{SHU})$. Curated $\rho = 0.76$, raw $\rho = 0.52$.}
\label{fig:gt_scoville}
\end{figure}

Table~\ref{tab:gt_ordinal} summarizes the ordinal projection results across all continuous and ordinal dimensions. \textbf{Scoville heat} shows the strongest ordinal signal ($\rho = 0.76$ curated, $p < .001$; Figure~\ref{fig:gt_scoville}), and the largest improvement from curation ($+0.24$). However, with only $n = 57$ spicy ingredients, the cross-validated estimate ($\rho_{\text{CV}} = 0.20$, $\sigma = 0.43$) shows substantial shrinkage; test folds of $\sim$6 ingredients produce inherently noisy Spearman estimates, so the true out-of-sample $\rho$ likely lies between these bounds. The heat axis in embedding space is aligned with perceived pungency from mild peppers through habanero and Thai chili.

\begin{figure}[!htbp]
\centering
\includegraphics[width=\textwidth]{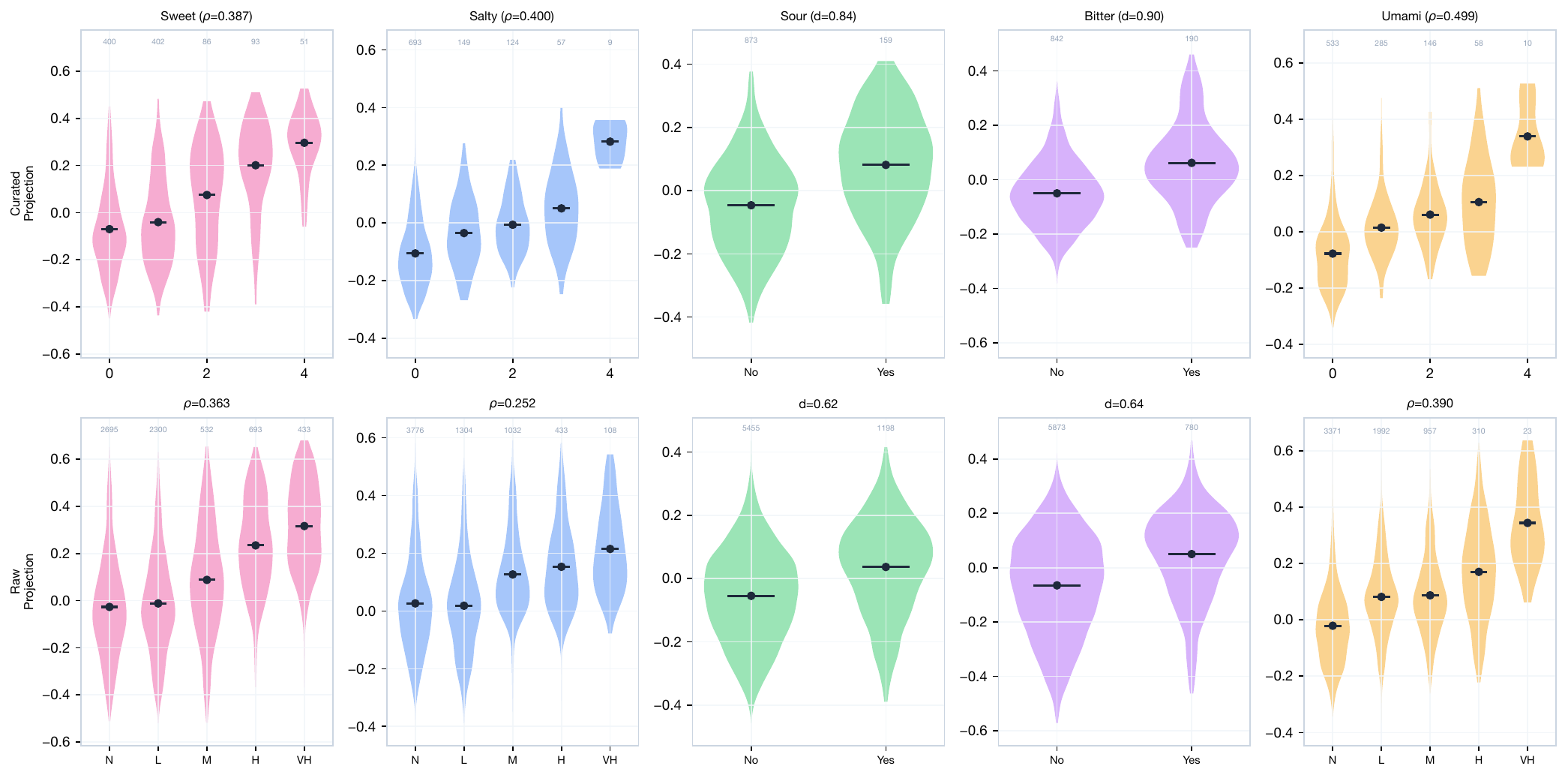}
\caption{Taste dimension violin plots. Top row: curated space (1,032 ingredients). Bottom row: raw space (6,653 ingredients). Sweet, salty, and umami use ordinal projection (5-level none$\to$very\_high axis, $\rho$ shown); sour and bitter use direct binary LLM labels (No/Yes, Cohen's $d$ shown). Numbers above violins show sample sizes.}
\label{fig:gt_taste_violins}
\end{figure}

\textbf{Umami} and \textbf{salty} exhibit robust monotonic gradients ($\rho = 0.50$ and $0.40$ respectively). Curation improves both substantially ($+0.11$ and $+0.15$), indicating that consolidating ingredient variants sharpens the savory dimension. \textbf{Sweet} shows a moderate correlation ($\rho = 0.39$) with modest curation improvement ($+0.02$), likely because sweet ingredients (fruits, sugars) were already well-separated in the raw space. \textbf{Sour} and \textbf{bitter} show near-zero ordinal correlation ($|\rho| < 0.1$), but as shown in Section~\ref{sec:extreme_results}, direct binary LLM classification reveals statistically significant embedding separation ($d = 0.84$ and $0.91$, both $p < .001$). Figure~\ref{fig:gt_taste_violins} shows the full projections.

\begin{figure}[!htbp]
\centering
\includegraphics[width=\textwidth]{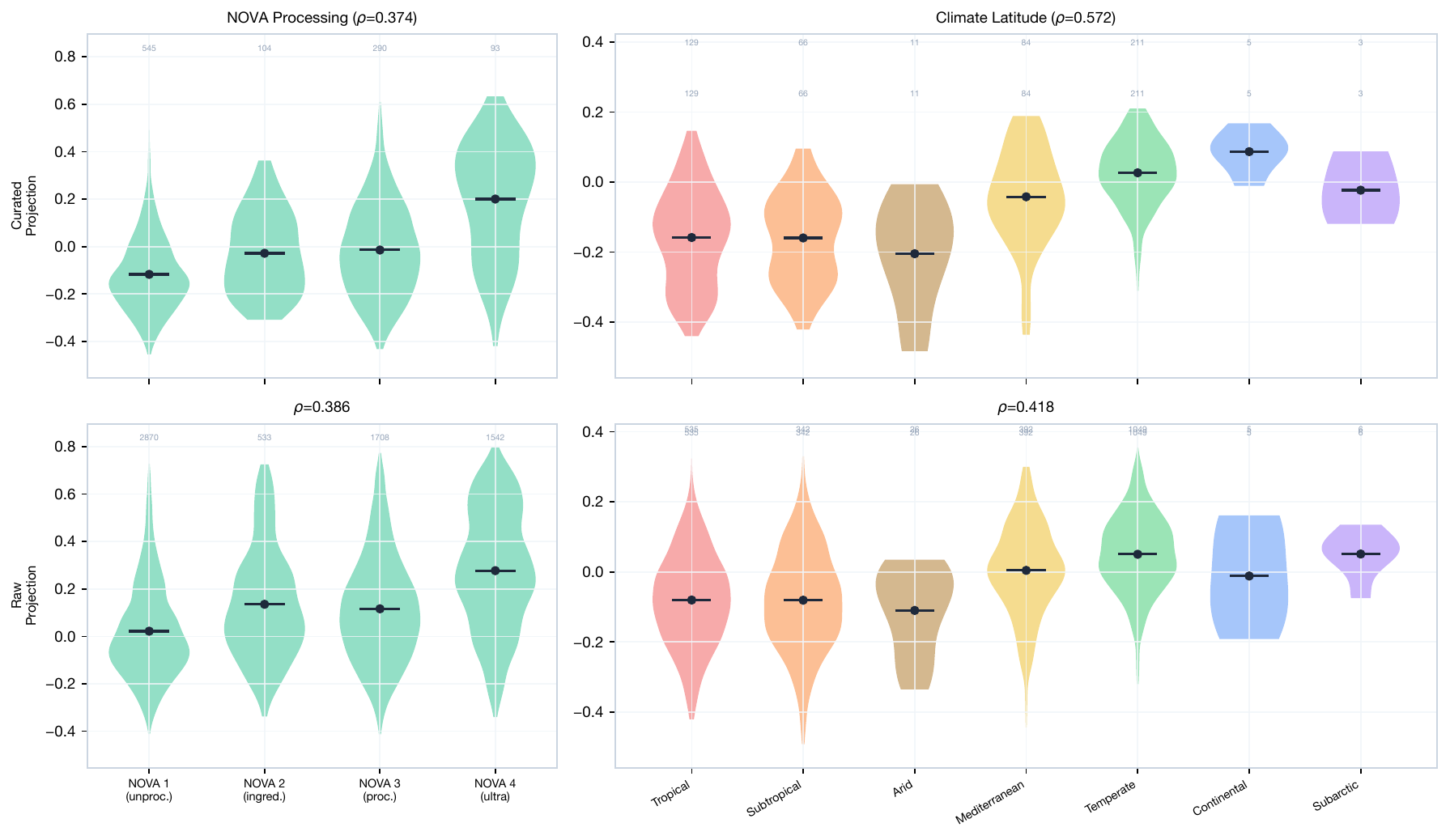}
\caption{Ordinal projection violin plots. \textbf{(a)}~NOVA processing level: the embedding axis separates unprocessed (NOVA~1) from ultra-processed (NOVA~4) ingredients (curated $\rho = 0.37$, raw $\rho = 0.39$). \textbf{(b)}~Climate latitude: projection onto the tropical$\to$subarctic axis, with per-zone violins ordered by latitude. Curated $\rho = 0.57$, raw $\rho = 0.42$. Top row: curated space. Bottom row: raw space. Axes are shared within each column.}
\label{fig:gt_nova_climate}
\end{figure}

\textbf{NOVA processing} correlates at $\rho = 0.37$ (curated) and $0.39$ (raw) (Figure~\ref{fig:gt_nova_climate}a), indicating that the embeddings encode the degree of food transformation, unprocessed ingredients project differently from ultra-processed products, but curation does not improve this dimension, possibly because processing state is orthogonal to the naming inconsistencies that curation addresses. Complementary evidence comes from 94 pre-specified raw$\to$processed ingredient pairs across 9 transformation categories (each pair links the same base ingredient before/after a defined processing operation): the mean cosine similarity is $0.541$ ($1.9\times$ random baseline), and the ranking is culinarily interpretable, identity-preserving operations (zesting, juicing) score highest ($\mu > 0.70$), while radical transformations (distilling, curing) score lowest ($\mu < 0.50$).

\textbf{Climate latitude} shows a strong signal ($\rho = 0.57$; Figure~\ref{fig:gt_nova_climate}b), with tropical ingredients (coconut, lemongrass, tamarind) projecting toward the low-latitude pole and temperate/continental ingredients (rye, barley, dill) toward the high-latitude pole. Curation improves this by $+0.15$.
\begin{table}[!htbp]
\centering
\caption{Ordinal projection analysis: Spearman $\rho$ between LLM-assigned rank and embedding projection onto the low-to-high axis. $\rho_{\text{CV}}$: 10-fold cross-validated $\rho$ (axis defined on training folds only; 20 random repeats). All $p < .001$ except sour ($p = 0.53$, curated) and bitter ($p = 0.42$, curated), which fail the ordinal test; these two dimensions are instead characterized via extreme-level analysis (Table~\ref{tab:extreme}). $n$: number of ingredients with valid tags.}
\label{tab:gt_ordinal}
\begin{tabular}{@{}lrcccc@{}}
\toprule
\textbf{Dimension} & $n_{\text{cur}}$ & \textbf{Curated $\rho$} & \textbf{$\rho_{\text{CV}}$} & \textbf{Raw $\rho$} & \textbf{$\Delta$} \\
\midrule
Scoville heat   &    57 & \cellcolor{hm9}0.757 & \cellcolor{hm6}0.20$^{\dagger}$ & \cellcolor{hm8}0.522 & \cellcolor{hm8}$+$0.235 \\
Climate latitude &   509 & \cellcolor{hm9}0.572 & \cellcolor{hm8}0.48 & \cellcolor{hm8}0.418 & \cellcolor{hm7}$+$0.154 \\
Umami            & 1,032 & \cellcolor{hm8}0.499 & \cellcolor{hm8}0.48 & \cellcolor{hm8}0.390 & \cellcolor{hm6}$+$0.110 \\
Salty            & 1,032 & \cellcolor{hm8}0.400 & \cellcolor{hm8}0.38 & \cellcolor{hm7}0.252 & \cellcolor{hm7}$+$0.148 \\
Sweet            & 1,032 & \cellcolor{hm8}0.387 & \cellcolor{hm8}0.37 & \cellcolor{hm8}0.363 & \cellcolor{hm5}$+$0.024 \\
NOVA processing  & 1,032 & \cellcolor{hm8}0.374 & \cellcolor{hm7}0.35 & \cellcolor{hm8}0.386 & \cellcolor{hm5}$-$0.012 \\
Bitter           & 1,032 & \cellcolor{hm5}0.025 & \cellcolor{hm5}0.01 & \cellcolor{hm6}0.093 & \cellcolor{hm5}noise \\
Sour             & 1,032 & \cellcolor{hm5}$-$0.020 & \cellcolor{hm5}$-$0.04 & \cellcolor{hm6}0.077 & \cellcolor{hm5}noise \\
\bottomrule
\multicolumn{6}{@{}l}{\footnotesize $^{\dagger}$High variance ($\sigma = 0.43$) due to small $n$; fold size $\approx 6$.}
\end{tabular}
\end{table}

\FloatBarrier
% ============================================================
\subsection{Extreme-Level Separation}
\label{sec:extreme_results}

Sour and bitter show near-zero ordinal $\rho$, indicating that the embeddings do not encode these tastes as linear gradients. We independently classify all ingredients with a direct binary question (``Is this ingredient notably sour/bitter? Yes/No'') using the same LLM (Gemini~3.1 Pro). The resulting binary axis, centroid of No-labeled vs.\ Yes-labeled ingredients, yields statistically significant embedding separation (Table~\ref{tab:extreme}, $p < .001$), confirming that the embeddings encode these taste dimensions as binary rather than graded signals.

\begin{table}[!htbp]
\centering
\caption{Binary separation analysis for taste dimensions. ``Direct binary $d$'': Cohen's $d$ from independent yes/no LLM labels. $d_{\text{CV}}$: 10-fold cross-validated $d$ (axis defined on training folds; 20 repeats). Ordinal $\rho$ from Table~\ref{tab:gt_ordinal} shown for comparison. All $p < .001$.}
\label{tab:extreme}
\scriptsize
\begin{tabular}{@{}lccccc@{}}
\toprule
& \multicolumn{3}{c}{\textbf{Direct binary $d$}} & \multicolumn{2}{c}{\textbf{Ordinal $\rho$}} \\
\cmidrule(lr){2-4}\cmidrule(lr){5-6}
\textbf{Taste} & \textbf{Cur.} & \textbf{$d_{\text{CV}}$} & \textbf{Raw} & \textbf{Cur.} & \textbf{Raw} \\
\midrule
Bitter  & \cellcolor{hm8}0.91 & \cellcolor{hm7}0.58 & \cellcolor{hm7}0.64 & 0.03 & 0.04 \\
Sour    & \cellcolor{hm8}0.84 & \cellcolor{hm7}0.58 & \cellcolor{hm7}0.62 & $-$0.02 & 0.00 \\
\bottomrule
\end{tabular}
\end{table}

Using direct binary labels, bitter achieves $d = 0.91$ and sour $d = 0.84$ despite having $\rho \approx 0$ ordinally. Cross-validated estimates ($d_{\text{CV}} = 0.58$ for both) confirm that the separation persists when the axis is defined on held-out data, with shrinkage of $0.26$--$0.33$ reflecting the in-sample inflation. All five tastes show statistically significant separation in both curated and raw spaces (all $p < .001$).

% ============================================================
\subsection{Taste Axis Geometry}
\label{sec:taste_geometry}

The five taste axes are not orthogonal. Sweet, salty, and umami axes are defined as the ordinal none$\to$very\_high centroid direction; sour and bitter axes use the binary No$\to$Yes centroid direction from direct LLM labels. Table~\ref{tab:taste_angles} shows the inter-axis cosine similarities.

\begin{table}[!htbp]
\centering
\caption{Inter-axis cosine similarity between the five basic taste axes in the curated 300d space. Sweet, salty, and umami use ordinal (none$\to$very\_high) axes; sour and bitter use binary (No$\to$Yes) axes. Values near 0 indicate orthogonal (independent) axes; values near $\pm 1$ indicate aligned (confounded) axes.}
\label{tab:taste_angles}
\begin{tabular}{@{}lccccc@{}}
\toprule
& \textbf{Sweet} & \textbf{Salty} & \textbf{Sour} & \textbf{Bitter} & \textbf{Umami} \\
\midrule
Sweet  & 1.00 & \cellcolor{hm1}$-$0.42 & \cellcolor{hm8}0.40 & \cellcolor{hm6}0.16 & \cellcolor{hm1}$-$0.45 \\
Salty  &      &  1.00   & \cellcolor{hm2}$-$0.33   & \cellcolor{hm3}$-$0.22 & \cellcolor{hm9}0.60 \\
Sour   &      &         &  1.00   & \cellcolor{hm5}0.01 & \cellcolor{hm2}$-$0.30 \\
Bitter &      &         &         &  1.00   & \cellcolor{hm2}$-$0.32 \\
Umami  &      &         &         &         &  1.00 \\
\bottomrule
\end{tabular}
\end{table}

The strongest coupling is between umami and salty ($\cos = 0.60$), consistent with the culinary overlap between these savory dimensions (soy sauce, miso, and aged cheeses are high in both) and with the broader multisensory flavor interactions reviewed by \cite{spence2020multisensory}. Sweet opposes both umami ($-0.45$) and salty ($-0.42$), reflecting the fundamental sweet--savory divide. With binary axes, sweet aligns strongly with sour ($0.40$), reflecting the pervasive sweet--sour pairing in fruits and condiments (citrus, vinegar-based sauces). Bitter is approximately orthogonal to sour ($0.01$) and weakly aligned with sweet ($0.16$), while opposing umami ($-0.32$), bitter ingredients (coffee, dark chocolate, hops) occupy a distinct niche from savory-umami ingredients.

\textbf{Partial correlations.} After regressing out projections on the other four taste axes, sweet retains the most unique signal (partial $\rho = 0.28$ vs.\ raw $\rho = 0.39$), while umami drops from $\rho = 0.50$ to partial $\rho = 0.15$, much of umami's apparent signal is shared with the salty axis.

\begin{figure}[!htbp]
\centering
\includegraphics[width=0.85\textwidth]{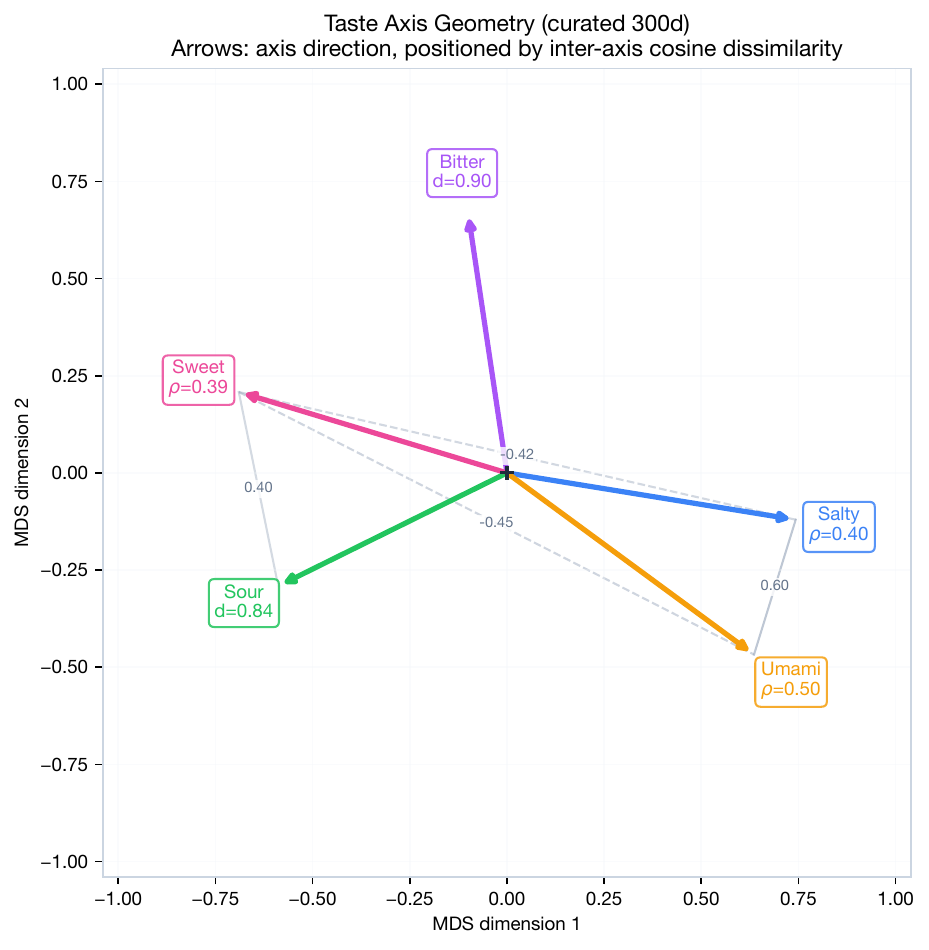}
\caption{Taste axis geometry via MDS projection of inter-axis cosine similarities. Sweet, salty, and umami use ordinal axes ($\rho$ shown); sour and bitter use direct binary axes (Cohen's $d$ shown). Solid lines indicate positive coupling ($\cos > 0.35$); dashed lines indicate negative coupling ($\cos < -0.35$).}
\label{fig:gt_taste_angles}
\end{figure}

\FloatBarrier
% ============================================================
\subsection{Texture Dimensions}
\label{sec:texture_results}

To test whether the embeddings encode textural properties, a modality distinct from taste and chemistry, we classify all ingredients along 6 texture dimensions grounded in the ISO~11036 sensory texture profile standard~\cite{iso11036} and the Szczesniak classification~\cite{szczesniak1963texture}: \emph{hardness} (force to compress), \emph{viscosity} (resistance to flow), \emph{crunchiness} (fracture behavior), \emph{chewiness} (mastication effort), \emph{moisture} (perceived water content), and \emph{fattiness} (perceived fat/oiliness). Each is initially tagged by Gemini~3.1 Pro on an ordinal scale (4--6 levels per dimension). Following the same rationale as for sour and bitter, five of the six dimensions are additionally classified with direct binary labels (``Is this ingredient hard/crunchy/chewy/moist/fatty? Yes/No''), with viscosity retaining ordinal analysis since it shows a robust gradient ($\rho = 0.43$).

Table~\ref{tab:texture_ordinal} summarizes all six texture dimensions. Five are characterized via binary separation (all $p < .001$), with curated Cohen's $d$ ranging from $0.79$ (chewiness) to $1.26$ (hardness). Viscosity is the only texture dimension with a robust ordinal gradient ($\rho = 0.43$). \textbf{Fattiness} shows the largest curation effect ($d$ from $0.48$ raw to $1.02$ curated), indicating that fat-perception encoding is obfuscated in the noisy raw space but emerges strongly after curation.

\begin{table}[!htbp]
\centering
\caption{Texture dimension analysis. \textbf{Left:} five binary dimensions, Cohen's $d$ from direct yes/no LLM labels; $d_{\text{CV}}$: 10-fold cross-validated (20 repeats). \textbf{Right:} viscosity (ordinal, restricted to liquids/semi-liquids). All $p < .001$.}
\label{tab:texture_ordinal}
\scriptsize
\begin{minipage}[t]{0.62\textwidth}
\centering
\begin{tabular}{@{}lcccc@{}}
\toprule
& \multicolumn{3}{c}{\textbf{Binary $d$}} & \\
\cmidrule(lr){2-4}
\textbf{Dimension} & \textbf{Cur.} & \textbf{$d_{\text{CV}}$} & \textbf{Raw} & \textbf{$\Delta d$} \\
\midrule
Hardness      & \cellcolor{hm8}1.26 & \cellcolor{hm8}0.84 & \cellcolor{hm7}0.59 & $+$0.67 \\
Moisture      & \cellcolor{hm8}1.17 & \cellcolor{hm8}1.00 & \cellcolor{hm7}0.63 & $+$0.54 \\
Fattiness     & \cellcolor{hm8}1.02 & \cellcolor{hm8}0.83 & \cellcolor{hm6}0.48 & $+$0.54 \\
Crunchiness   & \cellcolor{hm8}0.93 & \cellcolor{hm7}0.66 & \cellcolor{hm6}0.48 & $+$0.45 \\
Chewiness     & \cellcolor{hm7}0.79 & \cellcolor{hm7}0.60 & \cellcolor{hm6}0.52 & $+$0.27 \\
\bottomrule
\end{tabular}
\end{minipage}%
\hfill
\begin{minipage}[t]{0.35\textwidth}
\centering
\begin{tabular}{@{}lrcc@{}}
\toprule
& \multicolumn{3}{c}{\textbf{Ordinal $\rho$}} \\
\cmidrule(lr){2-4}
\textbf{Dim.} & $n$ & \textbf{Cur.} & \textbf{Raw} \\
\midrule
Viscosity & 285 & \cellcolor{hm8}0.43 & \cellcolor{hm8}0.38 \\
 & & \multicolumn{2}{c}{\footnotesize $\rho_{\text{CV}} = 0.35$} \\
\bottomrule
\end{tabular}
\end{minipage}
\end{table}

\begin{figure}[!htbp]
\centering
\includegraphics[width=\textwidth]{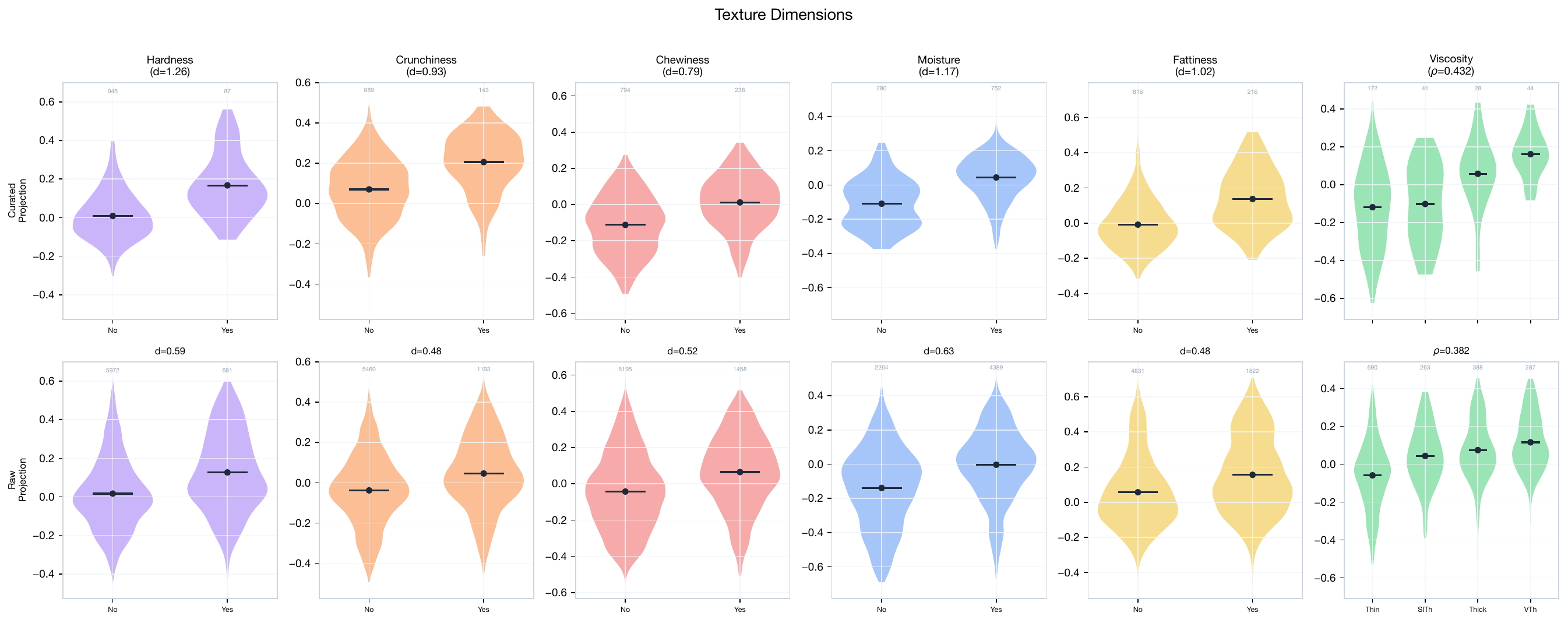}
\caption{Texture dimension violin plots. Top row: curated space (1,032 ingredients). Bottom row: raw space (6,653 ingredients). Viscosity uses ordinal projection (thin$\to$very\_thick axis, $\rho$ shown); the remaining five dimensions use direct binary LLM labels (No/Yes, Cohen's $d$ shown). Numbers above violins show sample sizes.}
\label{fig:gt_texture_violins}
\end{figure}

\begin{figure}[!htbp]
\centering
\includegraphics[width=0.85\textwidth]{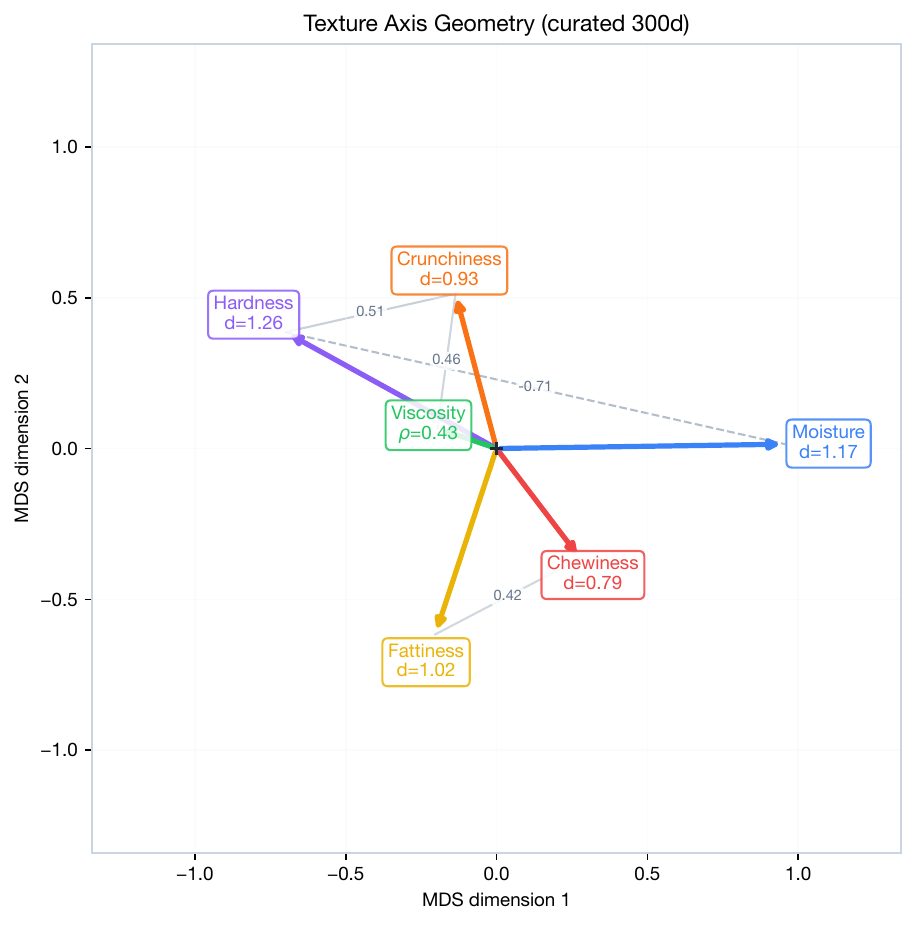}
\caption{Texture axis geometry via MDS. Five binary axes (No$\to$Yes direction, $d$ shown) plus viscosity ordinal axis ($\rho$ shown), positioned by inter-axis cosine dissimilarity. Solid/dashed lines indicate positive/negative coupling ($|\cos| > 0.3$).}
\label{fig:gt_texture_angles}
\end{figure}

\textbf{Texture axis geometry.} The six texture axes, five from binary labels, one ordinal, exhibit physically interpretable coupling (Figure~\ref{fig:gt_texture_angles}). Hardness and crunchiness are strongly correlated ($\cos = 0.51$): hard foods fracture audibly. Hardness opposes moisture ($\cos = -0.71$): dry foods are hard, wet foods are soft. Crunchiness aligns with viscosity ($\cos = 0.46$), and chewiness with fattiness ($\cos = 0.42$). These couplings mirror the known physics of food texture and confirm that the geometric relationships recovered from recipe co-occurrence are grounded in real physical constraints.

\FloatBarrier
% ============================================================
\subsection{Chemical and Nutritional Validation}
\label{sec:chemical_validation}

The preceding analyses use LLM-generated labels as proxy measurements. To test whether the recovered signals reflect genuine physicochemical structure, we cross-validate against laboratory-measured data from USDA FoodData Central and FooDB (methodology in Section~\ref{sec:chemical_methods}).

\subsubsection{Cross-Validation: LLM Axes vs.\ Measured Chemistry}

Table~\ref{tab:crossval} compares LLM-validated correlations with independent chemical measurements for four dimensions (three taste and moisture). The sweet axis shows the strongest chemical validation: both USDA total sugars ($\rho = 0.41$, $n = 416$) and FooDB total sugars (sucrose $+$ fructose $+$ glucose; $\rho = 0.40$, $n = 430$) track the embedding axis closely, with near-identical effect sizes from independent databases. Sodium--salty ($\rho = 0.29$, $n = 684$) is notably weaker, likely reflecting the complexity of perceived saltiness: many foods contain substantial sodium without tasting ``salty'' (bread, cheese, canned vegetables), and the umami--salty overlap in fermented products (soy sauce, miso) blurs the boundary between these perceptual dimensions. The glutamic acid--umami correlation ($\rho = 0.24$, $n = 494$) is modest but highly significant ($p < .001$). Moisture shows a strong signal ($\rho = 0.40$, $n = 712$), consistent with water content being a primary determinant of ingredient texture. Taken together, these correlations provide convergent evidence that embedding projections track meaningful physicochemical variation, while remaining associative rather than causal.

\begin{table}[!htbp]
\centering
\caption{Cross-validation of LLM-classified taste axes against measured chemical values. LLM~$\rho$: Spearman correlation from ordinal projection (Section~\ref{sec:gt_results}). Cur./Raw~$\rho$: correlation of the embedding axis with laboratory-measured values in the curated and raw spaces respectively. All correlations are significant ($p < .001$). Of the compounds measured here, only L-glutamic acid is a node in FlavorGraph's training graph; sucrose, glucose, fructose, sodium, and water are absent from the 1,561 training compounds (see Section~\ref{sec:chemical_methods}).}
\label{tab:crossval}
\begin{tabular}{@{}llrcrccc@{}}
\toprule
\textbf{Dimension} & \textbf{Chemical proxy} & \textbf{Source} & \textbf{LLM $\rho$} & $n$ & \textbf{Cur.\ $\rho$} & \textbf{Raw $\rho$} & \textbf{$\Delta$} \\
\midrule
Sweet   & Total sugars (g/100\,g)   & USDA  & 0.39 & 416 & \cellcolor{hm8}0.409 & 0.402 & \cellcolor{hm5}$+$0.01 \\
Sweet   & Total sugars (mg/100\,g)  & FooDB & 0.39 & 430 & \cellcolor{hm8}0.402 & 0.384 & \cellcolor{hm5}$+$0.02 \\
Salty   & Sodium (mg/100\,g)        & USDA  & 0.40 & 684 & \cellcolor{hm7}0.286 & 0.164 & \cellcolor{hm7}$+$0.12 \\
Umami   & L-Glutamic acid (mg/100\,g) & FooDB & 0.50 & 494 & \cellcolor{hm7}0.242 & 0.159 & \cellcolor{hm6}$+$0.08 \\
Moisture & Water (g/100\,g)         & USDA  & 0.38 & 712 & \cellcolor{hm8}0.397 & 0.292 & \cellcolor{hm7}$+$0.11 \\
\bottomrule
\end{tabular}
\end{table}

\subsubsection{Nutritional Dimensions}

We next test whether the embeddings encode macronutrient properties. For each nutrient, we define an axis via tercile centroids (bottom-third vs.\ top-third of measured USDA values) and project all matched ingredients, following the same methodology as the ordinal analyses above. No LLM is involved in axis definition or label assignment, these axes are defined entirely from laboratory measurements.

\begin{table}[!htbp]
\centering
\caption{Nutritional dimensions: Spearman $\rho$ between embedding projection (tercile-centroid axis) and USDA-measured nutritional values. Cur./Raw: curated (1,032) and raw (6,653) embedding spaces. $\rho_{\text{CV}}$: 10-fold cross-validated curated $\rho$ (axis defined on training folds, correlation measured on held-out fold; 20 random repeats). While aggregate macronutrient labels were not part of FlavorGraph's training data, some constituent molecules (12 amino acids, 35 fatty acids) are among the 1,561 training compounds; the hub ablation (Table~\ref{tab:hub_ablation}) quantifies the relative contribution of chemical vs.\ co-occurrence signal.}
\label{tab:nutritional}
\begin{tabular}{@{}lrcrrcc@{}}
\toprule
\textbf{Nutrient} & $n$ & \textbf{Cur.\ $\rho$} & \textbf{$\rho_{\text{CV}}$} & \textbf{$p$} & \textbf{Raw $\rho$} & \textbf{$\Delta$} \\
\midrule
Protein (g/100\,g)           & 712 & \cellcolor{hm8}0.473 & \cellcolor{hm8}0.40 & $6.8 \times 10^{-41}$ & 0.304 & \cellcolor{hm7}$+$0.17 \\
Carbohydrates (g/100\,g)     & 712 & \cellcolor{hm8}0.471 & \cellcolor{hm8}0.42 & $1.4 \times 10^{-40}$ & 0.339 & \cellcolor{hm7}$+$0.13 \\
Energy (kcal/100\,g)         & 712 & \cellcolor{hm8}0.469 & \cellcolor{hm8}0.40 & $2.7 \times 10^{-40}$ & 0.308 & \cellcolor{hm7}$+$0.16 \\
Dietary fiber (g/100\,g)     & 670 & \cellcolor{hm8}0.443 & \cellcolor{hm8}0.37 & $1.6 \times 10^{-33}$ & 0.272 & \cellcolor{hm7}$+$0.17 \\
Total fat (g/100\,g)         & 712 & \cellcolor{hm8}0.441 & \cellcolor{hm7}0.34 & $3.2 \times 10^{-35}$ & 0.214 & \cellcolor{hm8}$+$0.23 \\
\bottomrule
\end{tabular}
\end{table}

All five macronutrient dimensions yield highly significant correlations ($p < .001$), with effect sizes comparable to or exceeding many of the LLM-validated taste and texture dimensions. Protein ($\rho = 0.47$) and carbohydrates ($\rho = 0.47$) show the strongest signals. Cross-validated $\rho$ values range from $0.34$ (total fat) to $0.42$ (carbohydrates), representing a shrinkage of $0.06$--$0.10$ from full-data estimates. All cross-validated correlations remain substantial. The likely mechanism is that macronutrient composition constrains culinary function: high-protein ingredients serve as centerpieces, high-fat ingredients as cooking media, and high-carbohydrate ingredients as bases and binders. These functional roles determine recipe co-occurrence patterns, which the embeddings capture. The hub ablation below quantifies the extent to which each nutritional signal arises from chemical compound edges vs.\ co-occurrence patterns.

\begin{figure}[!htbp]
\centering
\includegraphics[width=\textwidth]{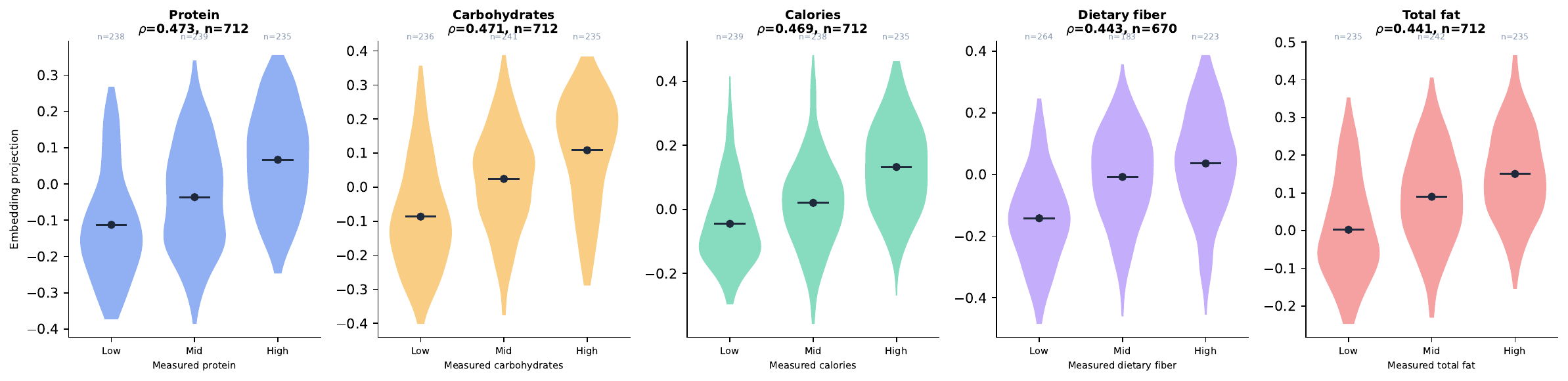}
\caption{Nutritional dimensions: violin plots of embedding projections for ingredients grouped by USDA-measured tercile (Low/Mid/High). Clear monotonic separation confirms that the embedding encodes each nutrient. All correlations are highly significant ($p < .001$).}
\label{fig:nutritional}
\end{figure}

\subsubsection{Chemical Hub Ablation}
\label{sec:hub_ablation}

Of the 1,032 curated ingredients, 308 are ``chemical hubs'', FlavorGraph nodes with direct compound edges in the training graph. The remaining 724 received chemical information only indirectly via metapath propagation through shared recipe contexts. The 1,561 FlavorDB flavor compounds that form these edges (35,440 edges total) are predominantly volatile aroma molecules; crucially, sucrose, glucose, fructose, sodium, and water are \emph{absent}, while 12 amino acids (189 edges, 0.5\%) and 35 fatty acids (1,289 edges, 3.6\%) are present. We report correlations separately for each subset across four dimensions with varying chemical provenance (Table~\ref{tab:hub_ablation}).

\begin{table}[!htbp]
\centering
\caption{Hub ablation: correlations for chemical-hub vs.\ non-hub ingredients in curated and raw spaces. Hub ingredients have direct compound edges in FlavorGraph's training graph; non-hub ingredients received chemical information only via metapath propagation. Total sugars and glutamic acid use FooDB compound measurements projected onto LLM-defined taste axes; protein and total fat use USDA nutrient measurements with tercile-centroid axes. The ``Training compounds'' column indicates the chemical provenance: whether the measured molecule(s) are among the 1,561 FlavorDB training compounds. $n$ and $p$ are for the curated space.}
\label{tab:hub_ablation}
\begin{tabular}{@{}llrcrccp{3.0cm}@{}}
\toprule
\textbf{Dimension} & \textbf{Subset} & $n$ & \textbf{Cur.\ $\rho$} & \textbf{$p$} & \textbf{Raw $\rho$} & \textbf{Training compounds} \\
\midrule
\multirow{2}{*}{Total sugars}  & Hub     & 156 & \cellcolor{hm8}0.461 & $1.4 \times 10^{-9}$ & 0.359 & \multirow{2}{3.0cm}{\footnotesize None (sucrose, glucose, fructose absent)} \\
                                & Non-hub & 274 & \cellcolor{hm8}0.373 & $1.9 \times 10^{-10}$ & 0.387 & \\
\midrule
\multirow{2}{*}{Glutamic acid} & Hub     & 194 & \cellcolor{hm7}0.305 & $1.5 \times 10^{-5}$ & 0.216 & \multirow{2}{3.0cm}{\footnotesize L-glutamic acid (1 compound)} \\
                                & Non-hub & 300 & \cellcolor{hm6}0.186 & $1.2 \times 10^{-3}$ & 0.158 & \\
\midrule
\multirow{2}{*}{Protein}       & Hub     & 279 & \cellcolor{hm8}0.466 & $2.0 \times 10^{-16}$ & 0.228 & \multirow{2}{3.0cm}{\footnotesize 12 amino acids (0.5\% of edges)} \\
                                & Non-hub & 433 & \cellcolor{hm8}0.475 & $9.2 \times 10^{-26}$ & 0.309 & \\
\midrule
\multirow{2}{*}{Total fat}     & Hub     & 279 & \cellcolor{hm8}0.506 & $1.4 \times 10^{-19}$ & 0.443 & \multirow{2}{3.0cm}{\footnotesize 35 fatty acids (3.6\% of edges)} \\
                                & Non-hub & 433 & \cellcolor{hm8}0.403 & $2.2 \times 10^{-18}$ & 0.201 & \\
\bottomrule
\end{tabular}
\end{table}

\textbf{Total sugars}: both hub ($\rho = 0.46$) and non-hub ($\rho = 0.37$) ingredients show highly significant correlations (both $p < .001$). Since none of these sugars are training compounds, the sweet signal arises entirely from recipe co-occurrence, sweet ingredients cluster in dessert, baking, and beverage recipes regardless of compound edges. \textbf{Glutamic acid}: hub ingredients show a stronger signal ($\rho = 0.31$) than non-hub ($\rho = 0.19$), consistent with L-glutamic acid being a training compound that provides a direct chemical graph pathway. \textbf{Protein}: hub ($\rho = 0.47$) and non-hub ($\rho = 0.48$) correlations are nearly identical despite 12 amino acids being training compounds, indicating that the protein signal arises primarily from co-occurrence. \textbf{Total fat}: the hub advantage ($\rho = 0.51$ vs.\ $0.40$) is consistent with 35 fatty acids providing a partial chemical pathway, yet non-hub correlation remains highly significant, confirming that co-occurrence also encodes fat content.

The pattern across dimensions is a gradient: some signals (sweet, salty, carbohydrates) arise primarily from co-occurrence with negligible chemical pathway, while others (umami, fat) have a measurable chemical component. Most are further strengthened by data curation (Section~\ref{sec:comparison}).

\FloatBarrier
% ============================================================
\subsection{Cultural Clustering}
\label{sec:cultural_results}

\subsubsection{LLM-Generated Distinctive Cultural Annotations}
\label{sec:cultural_clusters}

We test whether the embeddings encode cultural cuisine structure by examining whether ingredients from the same culinary tradition cluster in embedding space. We instruct Gemini~3.1 Pro to tag each of the 1,032 curated ingredients with cuisine associations \emph{only} if the ingredient is a \textbf{distinctive cultural marker}: one that uniquely signals a culinary tradition. Universal ingredients (garlic, chicken, butter, rice, onion) are left untagged.
We define 7 macro-regional clusters, each grouping cuisines with shared ingredient traditions:
\begin{itemize}
    \item \emph{Japanese}: solo, justified by Japan's island-geography culinary isolation and its uniquely distinctive fermentation/umami tradition (dashi, miso, wasabi, nori, bonito, yuzu, mochi)
    \item \emph{East Asian}: Chinese + Korean
    \item \emph{Southeast Asian}: Thai + Vietnamese + Filipino + Indonesian/Malaysian
    \item \emph{South Asian}: Indian + Pakistani + Sri Lankan + Bangladeshi
    \item \emph{Latin American}: Mexican + Brazilian + Peruvian + Caribbean
    \item \emph{Mediterranean}: Italian + French + Iberian + Greek + Levantine + North African
    \item \emph{Northern/Atlantic}: Scandinavian + British/Irish + German/Central European + Eastern European + American
\end{itemize}

This yields 522 distinctive-tagged ingredients, with 510 (49\%) classified as universal. Cluster sizes range from Mediterranean ($n_{\text{sav}}=106$) to Southeast Asian ($n_{\text{sav}}=24$) in the savoury subset. Spot-checking confirms expected behaviour: lemongrass $\to$ Southeast Asian; miso, wasabi, ponzu $\to$ Japanese; gochujang, gochugaru $\to$ East Asian; chipotle, tomatillo $\to$ Latin American; za'atar, harissa $\to$ Mediterranean; ranch dressing, BBQ sauce $\to$ Northern/Atlantic; while chicken, garlic, butter, rice, ginger, coconut, and onion are all untagged.

Because sweet cooking relies on largely universal ingredients across cultures, while cultural distinctiveness is concentrated in savoury traditions, we restrict all subsequent cuisine analysis to the \emph{savoury subset} (flavor profile = savoury): 589 of 1,032 curated ingredients and 2,916 of 4,804 raw ingredients.

\FloatBarrier
\subsubsection{Cultural Cluster Analysis in Native Embedding Space}
\label{sec:300d_analysis}

All quantitative claims about cultural clusters are grounded in the native 300-dimensional cosine-similarity embedding space. We apply identical analyses to the curated and raw spaces, using the same LLM-generated cuisine tags. For $k$NN purity comparisons, we additionally subsample the raw space to match the curated savoury pool size ($n = 589$) using 200 bootstrap iterations, controlling for the $5.0\times$ difference in pool size and the presence of duplicate ingredient representations in the raw space.

\textbf{Intra-cluster cosine similarity.} For each cuisine, we compute the mean pairwise cosine similarity among all its distinctive savoury members (Table~\ref{tab:intra_sim}). Every cuisine has higher intra-cluster similarity in the curated space (Wilcoxon signed-rank $W = 28$, $p = 0.008$, one-sided). The overall mean is $0.392$ (curated) vs.\ $0.315$ (raw), against a global savoury baseline of $0.295$ (curated) and $0.213$ (raw).

\begin{table}[!htbp]
\centering
\caption{Intra-cluster cosine similarity in the 300-dimensional embedding space (savoury ingredients only). Higher values indicate tighter cultural clusters. Curated exceeds raw for all 7 cuisines (Wilcoxon $W = 28$, $p = 0.008$). Sub: raw subsampled to $n=589$ (200 bootstrap iterations; 95\% CI shown).}
\label{tab:intra_sim}
\scriptsize
\begin{tabular}{@{}lrcccl@{}}
\toprule
\textbf{Cuisine} & $n_{\text{cur}}$ & \textbf{Curated} & \textbf{Raw} & \textbf{Raw (sub)} & \textbf{Sub 95\% CI} \\
\midrule
South Asian        &  27 & \cellcolor{hm9}0.502 & \cellcolor{hm7}0.380 & \cellcolor{hm7}0.386 & [0.293, 0.515] \\
East Asian         &  44 & \cellcolor{hm9}0.432 & \cellcolor{hm7}0.371 & \cellcolor{hm7}0.371 & [0.289, 0.485] \\
Southeast Asian    &  24 & \cellcolor{hm8}0.419 & \cellcolor{hm7}0.327 & \cellcolor{hm7}0.332 & [0.266, 0.416] \\
Latin American     &  44 & \cellcolor{hm8}0.400 & \cellcolor{hm6}0.313 & \cellcolor{hm6}0.313 & [0.264, 0.365] \\
Japanese           &  46 & \cellcolor{hm8}0.391 & \cellcolor{hm7}0.331 & \cellcolor{hm7}0.328 & [0.278, 0.379] \\
Mediterranean      & 106 & \cellcolor{hm6}0.288 & \cellcolor{hm6}0.242 & \cellcolor{hm6}0.242 & [0.222, 0.264] \\
Northern/Atlantic  &  62 & \cellcolor{hm6}0.310 & \cellcolor{hm5}0.241 & \cellcolor{hm5}0.244 & [0.214, 0.273] \\
\midrule
\textbf{Mean}      &     & \cellcolor{hm8}\textbf{0.392} & \cellcolor{hm6}0.315 & \cellcolor{hm6}0.317 & \\
\textbf{Global baseline} & & \cellcolor{hm6}0.295 & \cellcolor{hm5}0.213 & & \\
\bottomrule
\end{tabular}
\end{table}

\textbf{$k$NN purity.} For each tagged savoury ingredient, we find its $k=10$ nearest neighbours by cosine similarity in the full 300-dimensional space and compute the fraction sharing the same cuisine tag (Table~\ref{tab:knn_purity}). The curated space achieves mean purity $0.430$ vs.\ $0.306$ (raw) and $0.218$ (size-matched subsample). The curated-vs.-subsampled comparison is the fairest, controlling for pool size: Wilcoxon $W = 28$, $p = 0.008$.

\begin{table}[!htbp]
\centering
\caption{$k$NN purity ($k=10$) in the 300-dimensional embedding space (savoury). \emph{Purity}: fraction of same-cuisine neighbours. \emph{Lift}: purity divided by random baseline. Raw (sub): subsampled to $n=589$ (200 bootstrap iterations). Curated purity exceeds the size-matched raw baseline for 7/7 cuisines (Wilcoxon $p = 0.008$).}
\label{tab:knn_purity}
\scriptsize
\begin{tabular}{@{}lrcccccc@{}}
\toprule
& & \multicolumn{3}{c}{\textbf{Purity}} & \multicolumn{3}{c}{\textbf{Lift}} \\
\cmidrule(lr){3-5}\cmidrule(lr){6-8}
\textbf{Cuisine} & $n$ & \textbf{Cur.} & \textbf{Raw} & \textbf{Raw (sub)} & \textbf{Cur.} & \textbf{Raw} & \textbf{Raw (sub)} \\
\midrule
South Asian        &  27 & \cellcolor{hm9}0.589 & \cellcolor{hm8}0.376 & \cellcolor{hm7}0.227 & 12.8$\times$ & 21.9$\times$ & 13.4$\times$ \\
East Asian         &  44 & \cellcolor{hm9}0.466 & \cellcolor{hm7}0.299 & \cellcolor{hm7}0.237 & 6.2$\times$ & 11.9$\times$ & 9.7$\times$ \\
Japanese           &  46 & \cellcolor{hm9}0.539 & \cellcolor{hm8}0.412 & \cellcolor{hm7}0.277 & 6.9$\times$ & 14.0$\times$ & 9.4$\times$ \\
Latin American     &  44 & \cellcolor{hm8}0.475 & \cellcolor{hm7}0.316 & \cellcolor{hm6}0.236 & 6.4$\times$ & 8.2$\times$ & 6.1$\times$ \\
Mediterranean      & 106 & \cellcolor{hm8}0.458 & \cellcolor{hm7}0.389 & \cellcolor{hm6}0.308 & 2.5$\times$ & 2.9$\times$ & 2.3$\times$ \\
SE Asian           &  24 & \cellcolor{hm7}0.262 & \cellcolor{hm6}0.167 & \cellcolor{hm5}0.102 & 6.4$\times$ & 10.1$\times$ & 6.2$\times$ \\
Northern/Atlantic  &  62 & \cellcolor{hm6}0.219 & \cellcolor{hm6}0.182 & \cellcolor{hm5}0.140 & 2.1$\times$ & 2.5$\times$ & 2.0$\times$ \\
\midrule
\textbf{Mean}      &     & \cellcolor{hm8}\textbf{0.430} & \cellcolor{hm7}0.306 & \cellcolor{hm6}\textbf{0.218} & \textbf{6.2$\times$} & 10.2$\times$ & \textbf{7.0$\times$} \\
\bottomrule
\end{tabular}
\end{table}

The full raw space shows higher lift ($10.2\times$ vs.\ $6.2\times$), but this is inflated by the $5.0\times$ larger pool: with more ingredients, the random baseline drops, mechanically increasing lift. Subsampling produces lift of $7.0\times$, comparable to the curated value.

\textbf{Per-ingredient distance to centroid.} As a distribution-level test, we compute each tagged ingredient's cosine distance to its cuisine centroid. The curated space yields significantly smaller distances (mean $0.391$ vs.\ $0.469$; Mann-Whitney $U$, $p < .001$; Cohen's $d = 0.63$, a medium-to-large effect). This confirms that curation tightens cultural clusters at the level of individual ingredients, not just aggregate statistics.

\FloatBarrier
\subsubsection{Spatial Structure of Cuisine Clusters in 3D UMAP Space}
\label{sec:culinary_map}

\begin{figure}[!htbp]
\centering
\includegraphics[width=\textwidth]{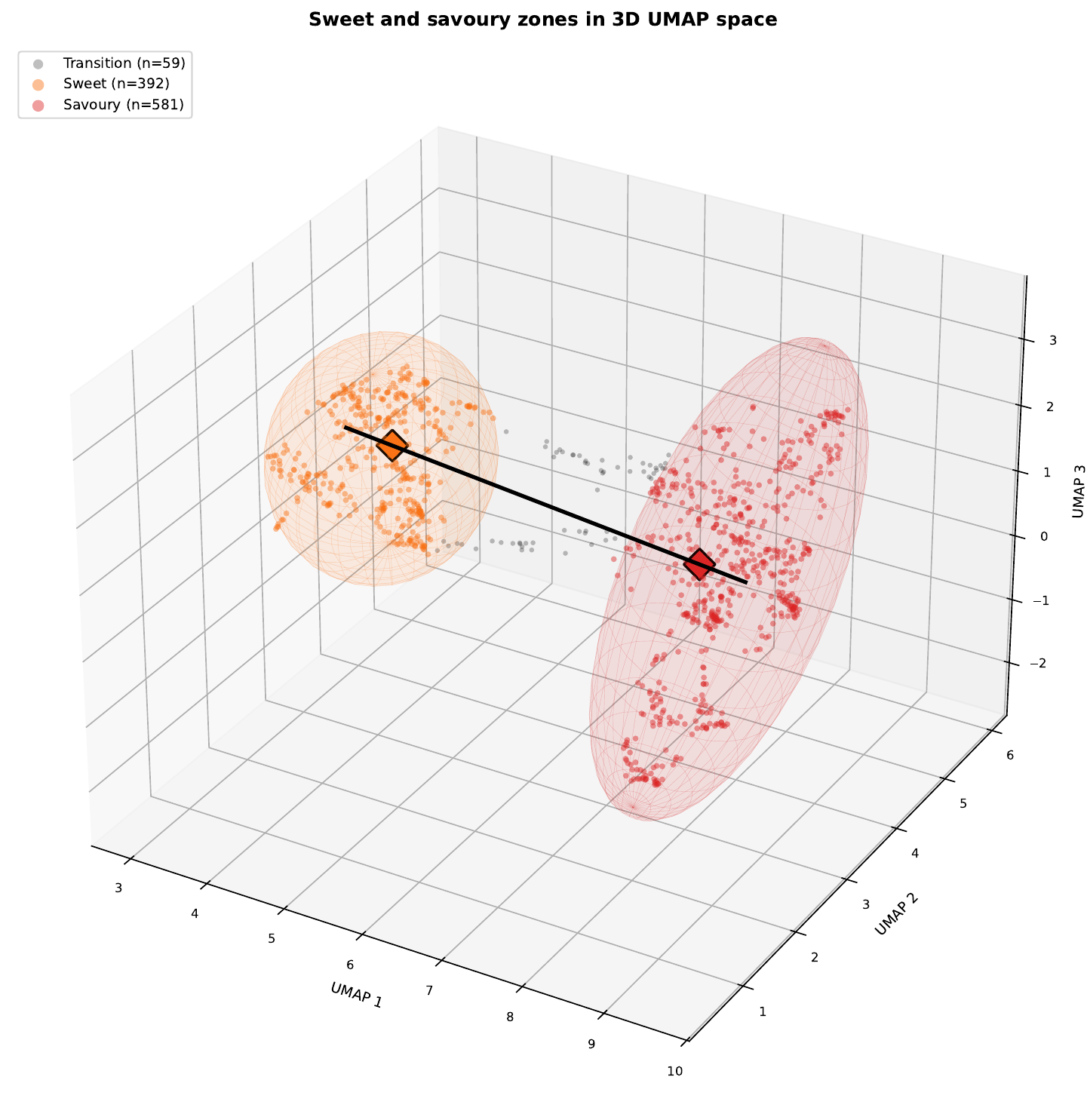}
\caption{3D UMAP projection of 1,032 curated ingredients. Sweet (orange) and savoury (red) zones are delineated by $2.5\sigma$ covariance ellipsoids around each pole centroid; transition-zone ingredients (black) lie between the poles. The black line marks the sweet$\leftrightarrow$savoury axis. An interactive 3D version is available at \url{https://epicure-data.kaikaku.ai}.}
\label{fig:cuisine_umap3d}
\end{figure}

\begin{figure}[!htbp]
\centering
\includegraphics[width=\textwidth]{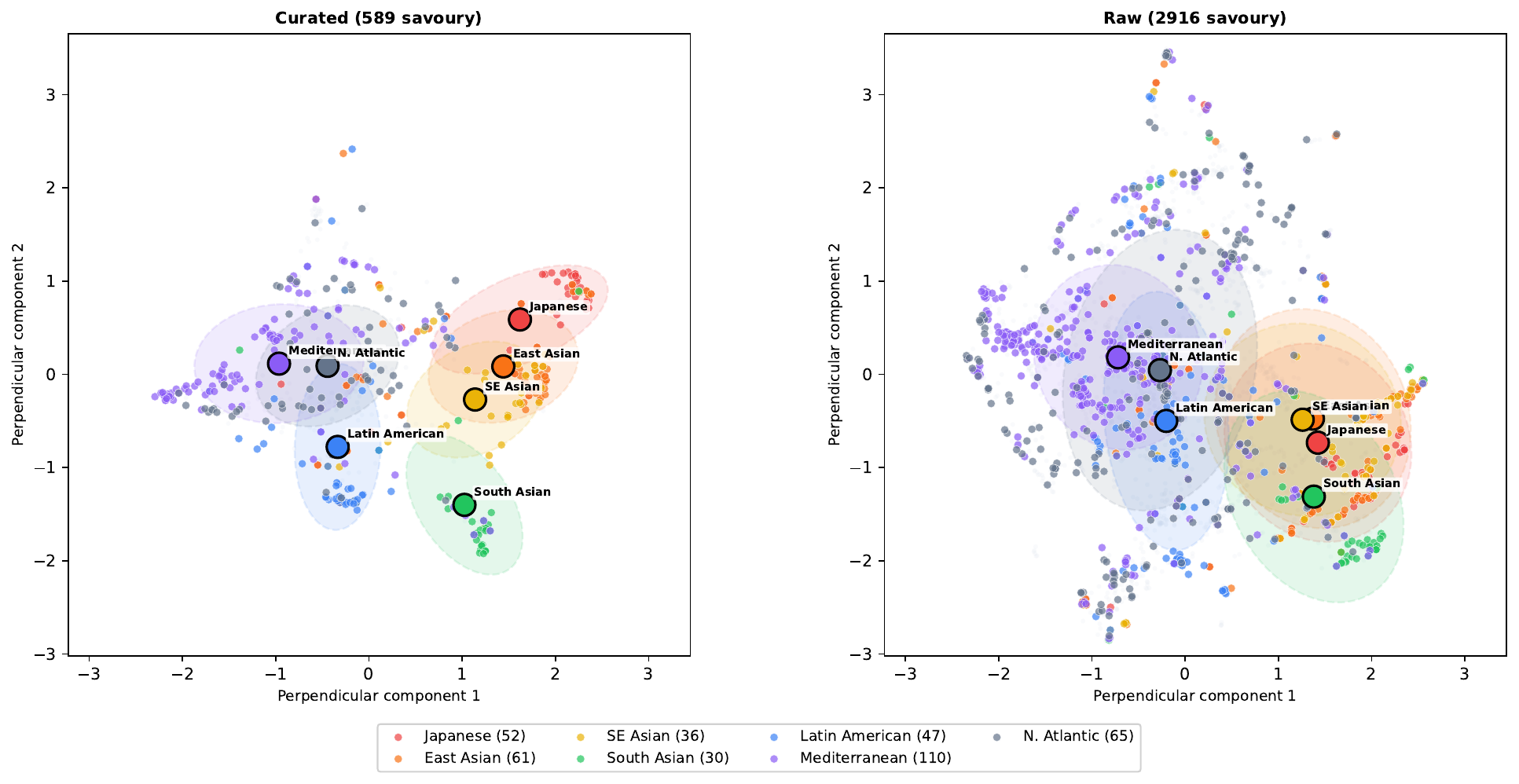}
\caption{Projection of savoury ingredients onto the plane perpendicular to the sweet$\leftrightarrow$savoury axis (Figure~\ref{fig:cuisine_umap3d}). Left: curated space. Right: raw space. Labelled circles mark cuisine centroids; dashed ellipses show $1\sigma$ covariance per cluster, visibly smaller in the curated space.}
\label{fig:perp_projection}
\end{figure}

The 300-dimensional analysis above establishes that cultural clusters exist and that curation sharpens them. The 3D UMAP projection (Appendix~\ref{sec:umap_toroid}) reveals a dominant sweet$\leftrightarrow$savoury axis separating two dense poles: the sweet pole ($n = 392$; beverages, sweets, fruits) and the savoury pole ($n = 581$; vegetables, meats, spices, condiments). We define this axis from the centroids of LLM-classified sweet and savoury ingredients, then project savoury ingredients onto the plane perpendicular to it (Figures~\ref{fig:cuisine_umap3d} and~\ref{fig:perp_projection}). The resulting 3D structure is best appreciated interactively at \url{https://epicure-data.kaikaku.ai}, revealing numerous substructures.

Figure~\ref{fig:cuisine_umap3d} shows the 3D UMAP projection with the sweet and savoury zones, the axis connecting their centroids, and $2.5\sigma$ covariance ellipsoids delineating each pole. We project savoury ingredients onto the plane perpendicular to the sweet--savoury axis at the savoury centroid (Figure~\ref{fig:perp_projection}): cuisines occupy distinct angular sectors, with Asian cuisines (Japanese, East Asian, Southeast Asian) separating from Mediterranean and Northern/Atlantic traditions. The $1\sigma$ covariance ellipses are visibly smaller in the curated space, corroborating the intra-cluster similarity gains reported in Table~\ref{tab:intra_sim}. The spatial separation on this projection plane confirms that the sweet--savoury axis is not the only organizing principle: within the savoury pole, cuisines are further differentiated by orthogonal dimensions that the embedding encodes.

\FloatBarrier
\subsubsection{Cuisine Profiles Across Classified Dimensions}
\label{sec:culture_dimensions}

The perpendicular projection above shows \emph{where} cuisines sit in the embedding space; we now quantify \emph{how} they differ. By projecting each cuisine centroid onto the dimension axes from Sections~\ref{sec:gt_results}--\ref{sec:taste_geometry}, we obtain a quantitative ``flavor profile'' per cuisine.

\begin{table}[!htbp]
\centering
\caption{Cuisine centroid projections onto classified dimension axes (curated 300d space). Each value is the cosine dot product of the L2-normalized cuisine centroid with the unit axis vector. Significance assessed by permutation test ($N = 10{,}000$; cuisine labels shuffled): ${}^{***}p < 0.001$, ${}^{**}p < 0.01$, ${}^{*}p < 0.05$; no star indicates non-significant.}
\label{tab:culture_profiles}
\scriptsize
\begin{tabular}{@{}lrrrrrrrr@{}}
\toprule
& \textbf{Umami$^{***}$} & \textbf{Salty$^{***}$} & \textbf{Sour$^{***}$} & \textbf{Scoville$^{***}$} & \textbf{Latitude$^{***}$} & \textbf{NOVA$^{**}$} & \textbf{Sweet} & \textbf{Bitter$^{***}$} \\
\midrule
Japanese       & \cellcolor{hm9}\textbf{0.45} & \cellcolor{hm6}0.09 & \cellcolor{hm3}$-$0.20 & \cellcolor{hm5}0.00 & \cellcolor{hm6}0.09 & \cellcolor{hm3}$-$0.15 & \cellcolor{hm3}$-$0.20 & \cellcolor{hm3}$-$0.18 \\
East Asian     & \cellcolor{hm7}0.24 & \cellcolor{hm6}0.13 & \cellcolor{hm2}$-$0.26 & \cellcolor{hm7}0.13 & \cellcolor{hm4}$-$0.12 & \cellcolor{hm3}$-$0.24 & \cellcolor{hm3}$-$0.25 & \cellcolor{hm4}$-$0.09 \\
SE Asian       & \cellcolor{hm7}0.14 & \cellcolor{hm6}0.12 & \cellcolor{hm3}$-$0.16 & \cellcolor{hm7}\textbf{0.20} & \cellcolor{hm3}$-$0.27 & \cellcolor{hm3}$-$0.24 & \cellcolor{hm3}$-$0.18 & \cellcolor{hm4}$-$0.07 \\
South Asian    & \cellcolor{hm5}$-$0.02 & \cellcolor{hm5}0.00 & \cellcolor{hm2}$-$0.24 & \cellcolor{hm5}0.01 & \cellcolor{hm1}\textbf{$-$0.35} & \cellcolor{hm2}\textbf{$-$0.30} & \cellcolor{hm3}$-$0.16 & \cellcolor{hm8}\textbf{0.31} \\
Latin American & \cellcolor{hm5}0.01 & \cellcolor{hm4}$-$0.11 & \cellcolor{hm5}0.01 & \cellcolor{hm5}$-$0.01 & \cellcolor{hm2}$-$0.30 & \cellcolor{hm2}$-$0.24 & \cellcolor{hm3}$-$0.21 & \cellcolor{hm4}$-$0.16 \\
Mediterranean  & \cellcolor{hm5}$-$0.05 & \cellcolor{hm5}$-$0.07 & \cellcolor{hm4}$-$0.04 & \cellcolor{hm5}0.00 & \cellcolor{hm6}0.05 & \cellcolor{hm3}$-$0.22 & \cellcolor{hm4}$-$0.16 & \cellcolor{hm4}$-$0.09 \\
N.\ Atlantic   & \cellcolor{hm6}0.07 & \cellcolor{hm5}$-$0.03 & \cellcolor{hm4}$-$0.14 & \cellcolor{hm4}$-$0.10 & \cellcolor{hm6}\textbf{0.12} & \cellcolor{hm5}$-$0.10 & \cellcolor{hm5}$-$0.13 & \cellcolor{hm3}$-$0.19 \\
\bottomrule
\end{tabular}

\medskip

\begin{tabular}{@{}lrrrrrr@{}}
\toprule
& \textbf{Hardness$^{***}$} & \textbf{Crunchiness$^{***}$} & \textbf{Chewiness$^{***}$} & \textbf{Moisture$^{***}$} & \textbf{Fattiness$^{***}$} & \textbf{Viscosity$^{***}$} \\
\midrule
Japanese       & \cellcolor{hm4}$-$0.13 & \cellcolor{hm8}0.34 & \cellcolor{hm5}0.02 & \cellcolor{hm7}\textbf{0.21} & \cellcolor{hm4}$-$0.09 & \cellcolor{hm7}0.23 \\
East Asian     & \cellcolor{hm5}$-$0.03 & \cellcolor{hm9}\textbf{0.41} & \cellcolor{hm5}0.00 & \cellcolor{hm7}0.19 & \cellcolor{hm3}$-$0.14 & \cellcolor{hm8}\textbf{0.33} \\
SE Asian       & \cellcolor{hm5}0.05 & \cellcolor{hm8}\textbf{0.40} & \cellcolor{hm5}$-$0.09 & \cellcolor{hm7}0.12 & \cellcolor{hm4}$-$0.15 & \cellcolor{hm7}0.27 \\
South Asian    & \cellcolor{hm9}\textbf{0.40} & \cellcolor{hm9}\textbf{0.41} & \cellcolor{hm4}$-$0.13 & \cellcolor{hm3}$-$0.23 & \cellcolor{hm4}$-$0.10 & \cellcolor{hm7}0.20 \\
Latin American & \cellcolor{hm6}0.05 & \cellcolor{hm8}0.29 & \cellcolor{hm4}$-$0.07 & \cellcolor{hm6}0.04 & \cellcolor{hm5}0.08 & \cellcolor{hm6}0.11 \\
Mediterranean  & \cellcolor{hm6}0.04 & \cellcolor{hm6}0.14 & \cellcolor{hm6}\textbf{0.08} & \cellcolor{hm5}0.00 & \cellcolor{hm8}0.23 & \cellcolor{hm5}$-$0.01 \\
N.\ Atlantic   & \cellcolor{hm4}$-$0.07 & \cellcolor{hm6}0.12 & \cellcolor{hm5}0.03 & \cellcolor{hm5}0.05 & \cellcolor{hm8}\textbf{0.30} & \cellcolor{hm5}0.06 \\
\bottomrule
\end{tabular}
\end{table}

\begin{figure}[!htbp]
\centering
\includegraphics[width=\textwidth]{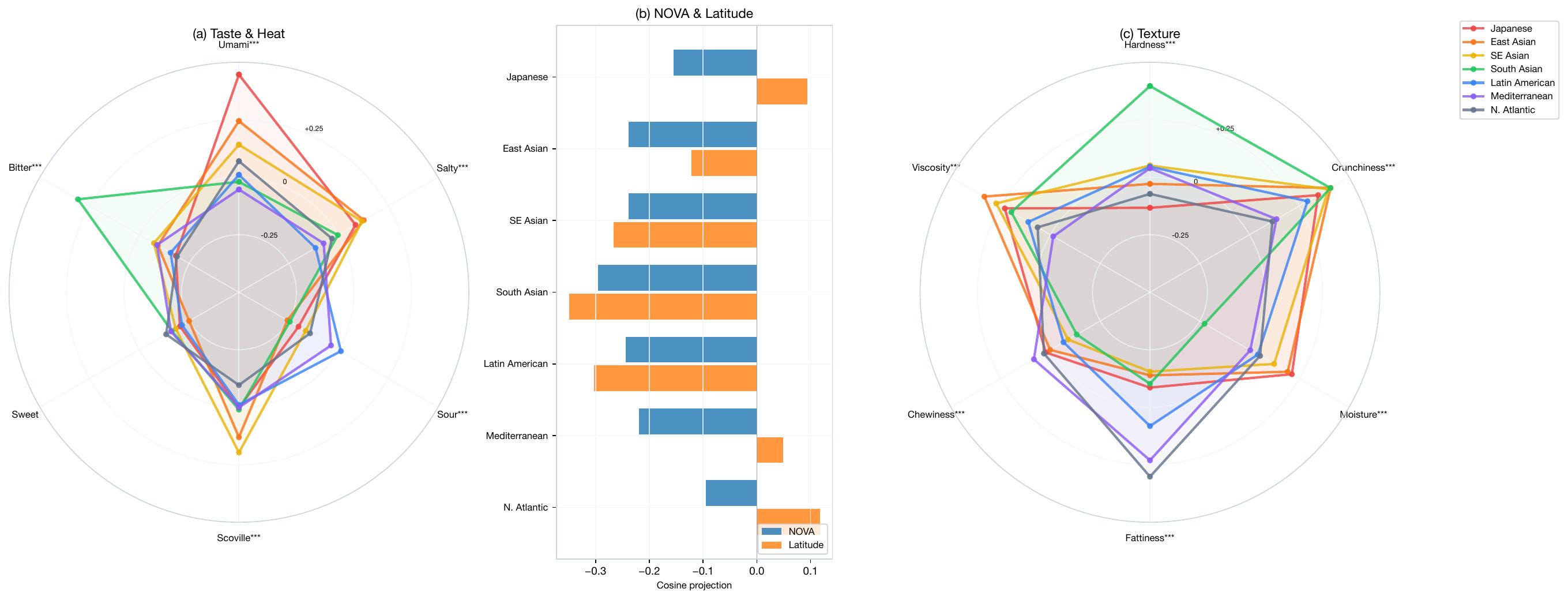}
\caption{Cuisine profiles across classified dimensions. (a)~Taste and heat dimensions (radar); (b)~NOVA and latitude (bar chart); (c)~Texture dimensions (radar). Radar panels show cosine projections shifted by $+0.5$ (concentric ring at 0.5 = zero). Significance stars from permutation test ($N=10{,}000$). 13 of 14 axes discriminate significantly between cuisines ($p < 0.01$ or better); Sweet is non-significant.}
\label{fig:culture_profiles}
\end{figure}

Table~\ref{tab:culture_profiles} and Figure~\ref{fig:culture_profiles} reveal interpretable cuisine signatures that align with culinary knowledge:

\begin{itemize}[noitemsep]
\item \textbf{Umami} separates Asian from Western cuisines, with Japanese dominating ($+0.45$), consistent with dashi, miso, soy, and bonito, while South Asian and Mediterranean cuisines are near-neutral.
\item \textbf{Scoville} peaks at Southeast Asian ($+0.20$) and East Asian ($+0.13$), reflecting the centrality of chili in Thai, Vietnamese, and Sichuan cooking, while Northern/Atlantic scores lowest ($-0.10$).
\item \textbf{Latitude} cleanly separates tropical-origin cuisines (South Asian $-0.35$, Latin American $-0.30$, SE Asian $-0.27$) from temperate/continental ones (N.\ Atlantic $+0.12$, Japanese $+0.09$).
\item \textbf{NOVA processing} shows all cuisines projecting toward the unprocessed pole; South Asian is most strongly unprocessed ($-0.30$), while Northern/Atlantic is least negative ($-0.10$), consistent with relatively higher use of processed condiments.
\item \textbf{Bitter} strongly distinguishes South Asian cuisine ($+0.31$), reflecting its reliance on inherently bitter ingredients (fenugreek, turmeric, mustard seed), while all other cuisines project toward the non-bitter pole.
\item \textbf{Sweet} is uniformly negative across all savoury-cuisine centroids, as expected, but this axis is non-significant in the cuisine-level permutation test.
\item \textbf{Crunchiness} is universally positive, all cuisines emphasise crunchy ingredients, with East Asian and South Asian highest (both $\approx +0.41$), consistent with legumes, seeds, and stir-fried vegetables.
\item \textbf{Fattiness} separates Western cuisines (N.\ Atlantic $+0.30$, Mediterranean $+0.23$; butter, cream, cheese, olive oil) from Asian cuisines, which project negatively.
\item \textbf{Hardness} peaks at South Asian ($+0.40$), whole spices and legumes, while Japanese ($-0.13$) and N.\ Atlantic ($-0.07$) are the softest.
\item \textbf{Moisture} separates Japanese ($+0.21$; seaweed, tofu, seafood) from the driest South Asian ($-0.23$; dry spices, lentils).
\end{itemize}

13 of 14 axes achieve significance in the permutation test (12 at $p < 0.001$, 1 at $p < 0.01$); Sweet is non-significant ($p = 0.135$), indicating that savoury-cuisine differentiation is not explained by sweetness variation.

\FloatBarrier
% ============================================================
\subsection{Raw vs.\ Curated Embedding Space Comparison}
\label{sec:comparison}

The curated space (1,032 ingredients) systematically outperforms the raw FlavorGraph space (4,804 ingredients with back-projected categories) on the majority of dimensions analyzed in Sections~\ref{sec:gt_results}--\ref{sec:texture_results}. The improvement arises from three interacting mechanisms: removal of non-food entries, population reweighting that eliminates the dominance of heavily-duplicated categories, and reduction of \emph{variant noise}, the intra-group scatter among entries that refer to the same canonical ingredient.

\subsubsection{Variant Noise: The Core Problem}

\begin{figure}[!htbp]
\centering
\includegraphics[width=\textwidth]{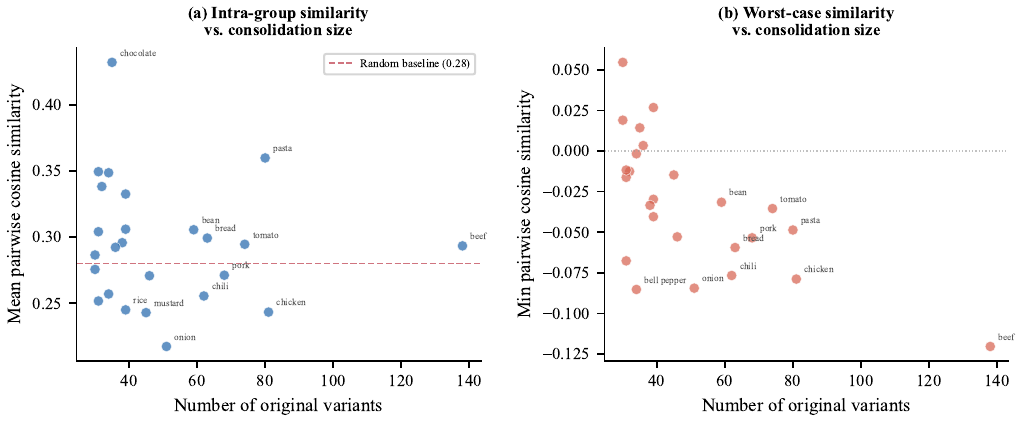}
\caption{Variant noise in the raw FlavorGraph data. (a)~Mean pairwise cosine similarity among all variants of a consolidated ingredient, vs.\ the number of variants. Most groups cluster near the random baseline (0.28), indicating that variants of the ``same'' ingredient are barely more similar to each other than to random ingredients. (b)~Minimum pairwise similarity: all 25 shown groups contain at least one pair with near-zero or negative cosine similarity.}
\label{fig:comparison_variants}
\end{figure}

To quantify the noise that curation removes, we examine the 50 most-consolidated ingredients (those with the most original FlavorGraph variants) and compute pairwise cosine similarity among all variants within each group (Figure~\ref{fig:comparison_variants}).

The results reveal severe intra-group incoherence: across the 50 groups, the mean of mean-pairwise-similarity is 0.31, barely above the random baseline of 0.28. All 50 groups have minimum pairwise similarity below 0.3, and all have at least one pair below 0.0 (orthogonal or opposing embeddings). The most extreme example is ``beef'' (138 variants, mean sim $= 0.29$, min sim $= -0.12$): entries like ``80\% lean ground beef'' and ``beef tenderloin tips'' occupy distant regions of embedding space because they co-occur in entirely different recipe contexts.

This scatter degrades raw-space statistics through three channels. First, each variant participates independently in every aggregate computation ($k$NN purity, Spearman $\rho$, Cohen's $d$), so heavily-duplicated categories dominate: beef's 138 entries contribute 138 votes to any measurement, while cumin contributes one. Second, many individual variants land in neighborhoods belonging to unrelated categories, inflating cross-category similarity and directly degrading $k$NN purity. Third, dimension axes defined from tercile centroids inherit the noise, because the ``high umami'' or ``high protein'' groups are populated by hundreds of scattered meat and cheese variants rather than unique concepts.

Consolidation via averaging (Equation~\ref{eq:embedding_avg}) addresses all three channels: it replaces each cluster of variants with a single representative, giving every canonical ingredient equal weight in aggregate statistics, removing cross-category contamination from variant scatter, and producing cleaner tercile centroids for axis definitions. Note that the GNN was trained on the noisy vocabulary from the start---there is no latent clean signal being ``restored.'' Rather, consolidation produces a measurement context in which the structure the GNN \emph{did} learn becomes easier to detect.

\begin{figure}[!htbp]
\centering
\includegraphics[width=\textwidth]{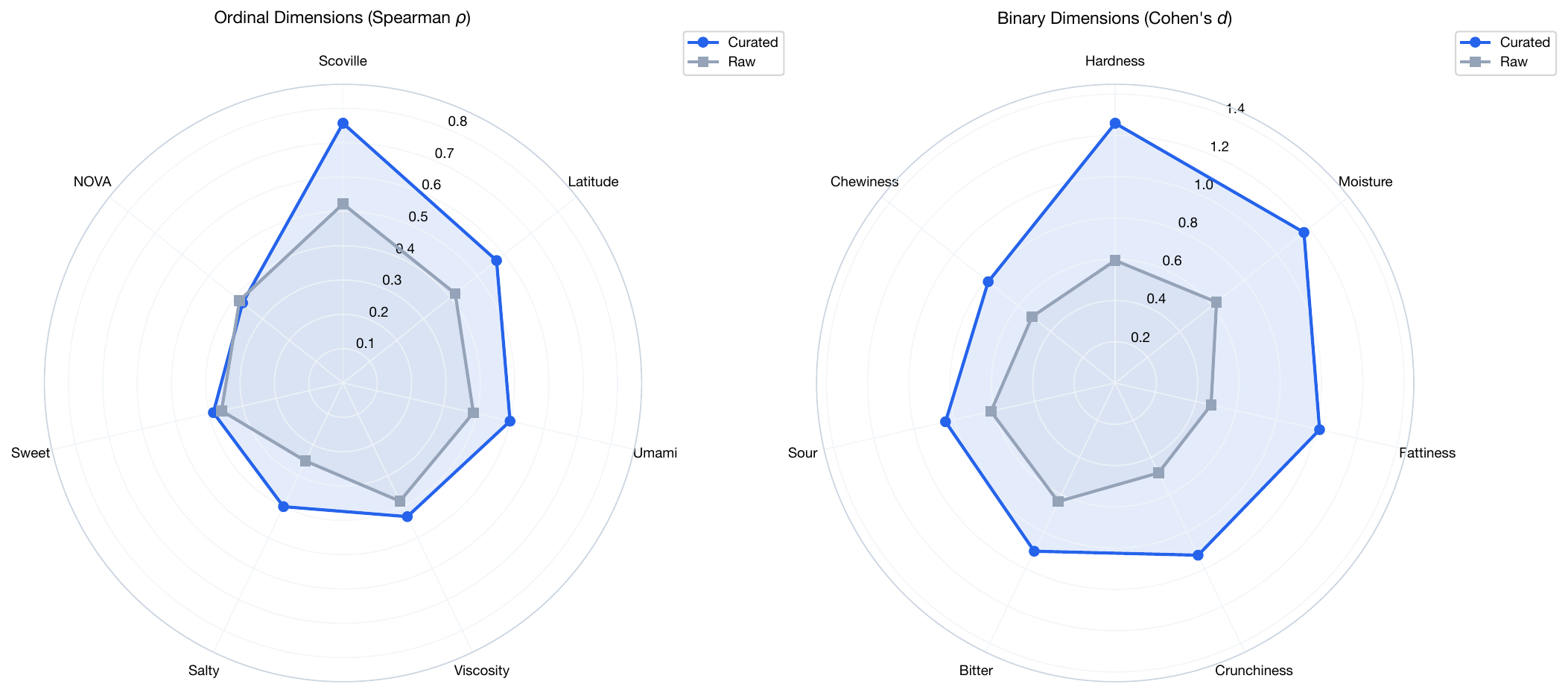}
\caption{Overview of curated vs.\ raw embedding space performance across all 14 validated dimensions. Left: ordinal dimensions (Spearman $|\rho|$), sweet, salty, umami, scoville, viscosity, NOVA, latitude. Right: binary dimensions (Cohen's $d$), sour, bitter, hardness, crunchiness, chewiness, moisture, fattiness. The curated space outperforms the raw space on the majority of dimensions.}
\label{fig:gt_radar}
\end{figure}

\begin{figure}[!htbp]
\centering
\includegraphics[width=\textwidth]{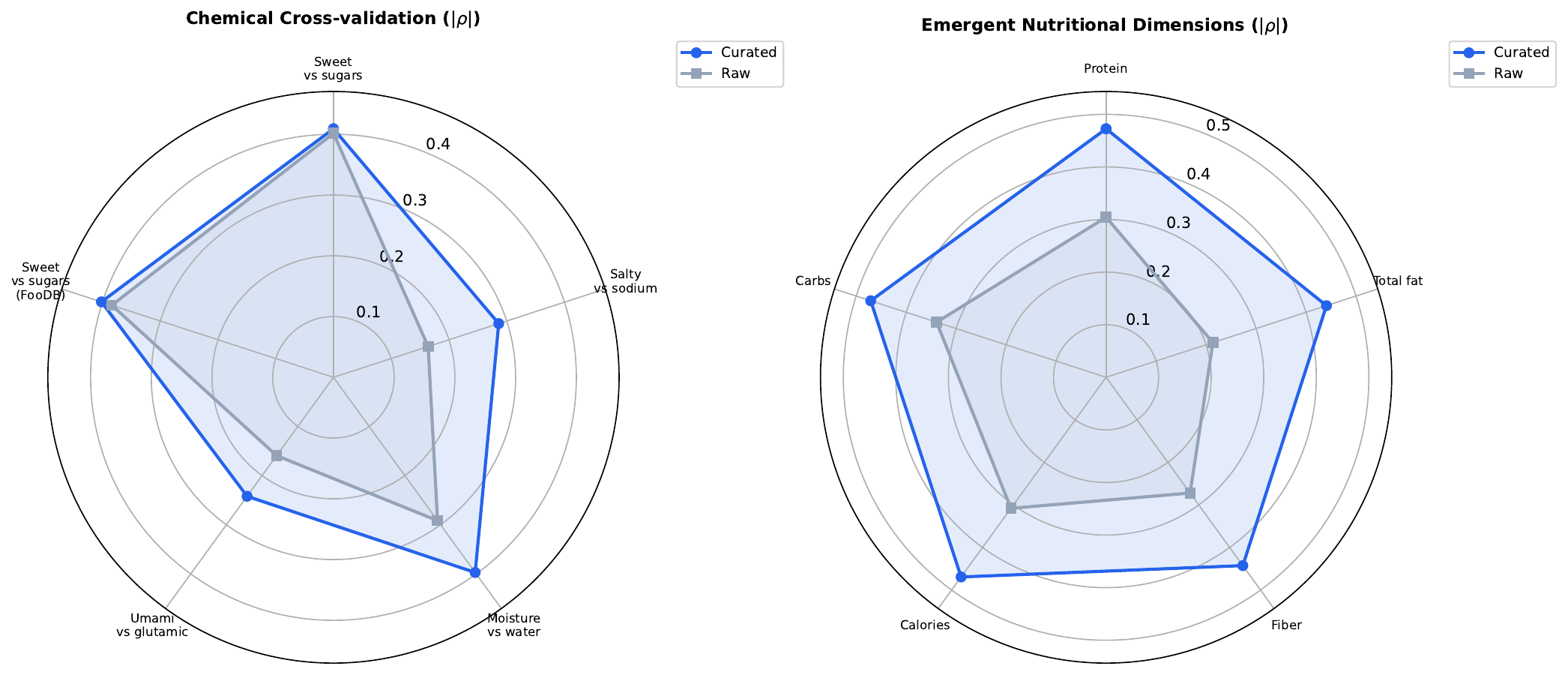}
\caption{Curated vs.\ raw embedding performance on external laboratory measurements. Left: chemical cross-validation (Spearman $|\rho|$) against USDA nutrients and FooDB compounds, the two embeddings perform similarly on sweetness, but the curated space yields substantially higher correlations for salty--sodium, moisture--water, and umami--glutamic acid. Right: nutritional dimensions from USDA tercile centroids, the curated space consistently outperforms raw across all five macronutrients (protein, fat, fiber, calories, carbohydrates). Raw ingredients are matched independently to USDA (5,430/6,653, 81.6\%) and FooDB (4,587/6,653, 68.9\%) using the same three-layer pipeline.}
\label{fig:nutritional_comparison}
\end{figure}

\FloatBarrier
% ============================================================
\section{Discussion}
\label{sec:discussion}

\subsection{Chemical Provenance and Signal Origins}

FlavorGraph embeddings encode culinary knowledge that arises from two intertwined sources: chemical compound edges and recipe co-occurrence patterns. The hub ablation (Section~\ref{sec:hub_ablation}) and our characterization of the 1,561 training compounds allow us to trace the provenance of each validated dimension.

The sweet--sugar correlation ($\rho = 0.41$) is especially informative: sucrose, glucose, and fructose are \emph{absent} from the training compounds, yet both hub ($\rho = 0.46$) and non-hub ($\rho = 0.37$) ingredients show strong correlations. Sweetness encoding arises from recipe co-occurrence, sweet ingredients cluster in dessert, baking, and beverage recipes. By contrast, umami encoding has a measurable chemical component: L-glutamic acid is a training compound, and hub ingredients show stronger correlations ($\rho = 0.31$) than non-hub ($\rho = 0.19$). Protein and fat sit between these poles: their constituent molecules (amino acids, fatty acids) are partially represented in the training graph, but the non-hub signal confirms that co-occurrence also contributes. Carbohydrates, fiber, and the taste, texture, cultural, and geographic dimensions have no or negligible representation among the training compounds.

This gradient is itself informative: recipe co-occurrence encodes culinary function, high-protein ingredients serve as centerpieces, high-fat ingredients as cooking media, sweet ingredients cluster in desserts, and these functional roles track physicochemical properties regardless of whether the relevant molecules appear in the training graph. The nutritional validation is particularly compelling because it involves no LLM at any stage: axes are defined from USDA tercile centroids, values are laboratory measurements, and correlations ($\rho = 0.34$--$0.42$ cross-validated, all $p < .001$) are comparable to or exceed many of the LLM-validated taste dimensions.

\subsection{Practical Applications}

These recovered dimensions are not merely scientifically interesting, they are directly actionable for food-technology systems. Because each dimension has a concrete direction in the 300-dimensional space, ingredient search, substitution, and recipe generation can operate on interpretable axes rather than opaque similarity scores.

\begin{itemize}[leftmargin=*,noitemsep]
\item \textbf{Multi-dimensional substitution.} Replacing an ingredient in a dish requires matching not one but several dimensions simultaneously. A vegan substitute for parmesan must score high on umami ($\rho = 0.50$), fattiness ($\rho = 0.22$), and hardness ($\rho = 0.23$), the embedding space enables nearest-neighbour search constrained to the relevant axes, yielding substitutes that preserve the \emph{functional role} in the dish rather than a single flavor note.

\item \textbf{Dietary accommodation by design.} Allergen and restriction constraints can be handled proactively: the system identifies alternatives that maintain the dimensional profile (fattiness, moisture, cultural cluster membership) while satisfying dietary requirements, enabling menus designed from the ground up for inclusivity.

\item \textbf{Quantitative cuisine fingerprints.} The cuisine centroid projections (Table~\ref{tab:culture_profiles}) provide a computational definition of each cuisine's identity across 14 axes. This enables recipe validation (``does this dish's ingredient set project onto the expected Japanese profile?'') and data-driven menu engineering for target cuisines or fusion concepts.

\item \textbf{Texture-aware recipe balancing.} Existing flavor-pairing tools ignore texture. The moisture--viscosity anti-correlation ($\cos = -0.66$) and hardness--chewiness coupling ($\cos = 0.57$) enable a system that ensures textural \emph{contrast} within a dish, pairing a creamy element with a crunchy one, or a wet sauce with a chewy protein, grounded in the physics of food rather than ad hoc rules.

\item \textbf{Cross-cultural fusion.} The perpendicular projection (Figure~\ref{fig:perp_projection}) reveals which cuisine clusters are geometrically adjacent, sharing compatible ingredient palettes. Japanese--Mediterranean fusion, for example, is supported by overlapping sour and umami profiles, while latitude separation supplies complementary diversity. These adjacencies can be explored in the interactive embedding viewer.\footnote{\url{https://epicure-data.kaikaku.ai}}

\item \textbf{Clean-label reformulation.} The NOVA processing dimension ($\rho = 0.37$) means the embedding space encodes the degree of ingredient transformation. Product developers can search for lower-processing alternatives that achieve equivalent taste and texture projections, replacing ultra-processed ingredients with minimally processed ones at matched sensory profiles.

\end{itemize}

Together, these applications illustrate how making tacit culinary knowledge computationally explicit can augment the creative processes of menu development, recipe innovation, and product formulation that the food-service and CPG industries currently perform through manual expertise alone.

\subsection{Current Findings as a Lower Bound}

Curation improves the dimensional correlation on the majority of dimensions tested, with improvements ranging from $+0.02$ (sweet) to $+0.24$ (Scoville) for taste/chemistry and up to $+0.21$ (fattiness) for texture dimensions. The raw embeddings already contain signal; curation makes it easier to detect by removing duplicate entries, rebalancing the population, and reducing the variant noise documented in Section~\ref{sec:comparison}. This has an important implication: the structure we measure in the current embeddings is a \emph{lower bound} on what these embeddings can encode. Better data, cleaner ingredient vocabularies, more diverse recipe corpora, richer graph structures, should yield stronger correlations. The texture results are particularly striking: viscosity ($\rho = 0.43$) and moisture ($\rho = 0.38$) are encoded as strongly as several taste dimensions. This suggests that recipe co-occurrence encodes functional roles that integrate across sensory modalities, as chefs obviously select ingredients not just for taste but for the textural balance they contribute to a dish.

\subsection{Future Directions}

The dimensions we measure are entangled in a single 300-dimensional vector, with chemical and co-occurrence signals inseparably mixed. Two complementary improvements could disentangle them: (1)~a high-quality relational database in which LLMs process raw recipe data into structured form, capturing processing chains, cultural contexts, and sensory profiles, moving beyond co-occurrence statistics; and (2)~multi-relational graph architectures that explicitly model distinct edge types (co-occurrence, chemical similarity, cultural tradition) and produce disentangled representations rather than conflating all dimensions into a single vector space. The taste axis geometry (Section~\ref{sec:taste_geometry}), where umami--salty coupling ($\cos = 0.60$) suppresses the sour signal, illustrates why such disentanglement matters. The hub ablation results show that the chemical and co-occurrence signals reinforce each other for some dimensions (sweetness) while remaining separable for others (umami), suggesting that disentanglement is feasible with appropriate architectures.

\subsection{Limitations}

\begin{itemize}[leftmargin=*]
\item \textbf{Static embeddings.} FlavorGraph embeddings are pre-computed and fixed. Embedding averaging is a lossy operation. Retraining on the cleaned vocabulary would avoid this information loss but requires access to the original Recipe1M+~\cite{marin2021recipe1m} corpus.

\item \textbf{Chemical training confound.} FlavorGraph's training data includes 1,561 flavor molecules sourced from FlavorDB~\cite{park2021flavorgraph,garg2018flavordb}, predominantly volatile aroma compounds, plus 12 amino acids (0.5\% of compound edges) and 35 fatty acids (3.6\%). Notably, sucrose, glucose, fructose, sodium, and water are absent from these training compounds. This means the sweet, salty, and moisture signals arise primarily from recipe co-occurrence, while protein and fat correlations have a partial chemical pathway through amino acid and fatty acid edges. We trace this provenance through the hub ablation analysis (Section~\ref{sec:hub_ablation}).

\item \textbf{LLM classifications are proxy labels, not ground truth.} The LLM-assigned classifications reflect Gemini's training data rather than measured sensory or chemical properties. Since both FlavorGraph and the LLM were trained on overlapping internet text (recipe websites, food science literature), there is a risk of information leakage. We mitigate this with USDA-based chemical cross-validation (Section~\ref{sec:chemical_validation}), which involves no LLM in axis definition or label assignment. Further validation with human sensory panels or direct chemical measurements would strengthen the analysis.

\item \textbf{Ordinal projection assumes linearity.} The ordinal projection method assumes that taste intensity varies linearly along a single axis in embedding space. The direct binary separation results (Section~\ref{sec:extreme_results}) demonstrate that this assumption is violated for sour and bitter, where encoding is non-linear.

\item \textbf{Binary granularity.} Our binary analyses use direct yes/no labels, which collapse intensity variation within the ``yes'' group. Larger ingredient vocabularies would enable finer-grained binning.

\item \textbf{English-centric vocabulary.} The 1,032-ingredient vocabulary is English-centric and biased toward ingredients common in Western recipe databases (the Recipe1M+~\cite{marin2021recipe1m} corpus is built from predominantly English-language recipe websites, many of them US-focused).

\item \textbf{Seven-cuisine granularity.} The 7 macro-clusters necessarily simplify the world's culinary diversity. Finer-grained clusters (e.g., separating Thai from Vietnamese) may not have sufficient distinctive members in this dataset for reliable analysis.

\item \textbf{UMAP topology.} The toroidal manifold structure (Appendix~\ref{sec:umap_toroid}) was selected from a grid search over 16 UMAP configurations (Appendix~\ref{sec:umap_params}), and the two-pole sweet--savoury structure also appears under t-SNE, though less distinctly. However, UMAP can produce artificial topological features that do not exist in the original high-dimensional space, and the finer toroidal geometry (transition zone, bridge ingredients) has not been verified under alternative methods.

\item \textbf{Axis-definition overfitting.} The ordinal and tercile-centroid axes are defined on the same data used to measure correlation, which inflates reported effect sizes. We report 10-fold cross-validated estimates for all dimensions (Tables~\ref{tab:gt_ordinal}--\ref{tab:nutritional}). For the large-$n$ ordinal dimensions ($n = 1{,}032$), shrinkage is small (0.02--0.02 for taste, 0.02 for NOVA). For medium-$n$ dimensions, shrinkage is moderate (latitude: $0.57 \to 0.48$; viscosity: $0.43 \to 0.35$). Scoville ($n = 57$) shows large shrinkage ($0.76 \to 0.20$), though this is partly an artifact of high variance in Spearman $\rho$ computed on $\sim$6-ingredient test folds. Binary dimensions show shrinkage of $0.17$--$0.42$ in Cohen's $d$, with all cross-validated estimates remaining substantial ($d_{\text{CV}} = 0.58$--$1.00$). Nutritional dimensions shrink by $0.06$--$0.10$ ($\rho_{\text{CV}} = 0.34$--$0.42$).
\end{itemize}

% ============================================================
\section{Conclusion}
\label{sec:conclusion}

FlavorGraph embeddings encode at least fifteen independently classifiable dimensions of culinary knowledge, spanning taste, texture, nutrition, geography, culture, and food processing. All survive ten-fold cross-validation, and the nutritional dimensions validate against USDA laboratory measurements with no LLM involvement at any stage.

The most instructive finding is what is \emph{not} in the training data. Sucrose, glucose, fructose, sodium, and water are absent from FlavorGraph's 1,561 training compounds, yet the embeddings encode sweetness, saltiness, and moisture as strongly as dimensions with direct chemical pathways. These signals arise entirely from recipe co-occurrence: sweet ingredients cluster in dessert and baking contexts, salty ingredients in savoury preparations, and the GNN recovers this structure without ever seeing the relevant molecules. The hub ablation traces a continuous gradient from pure co-occurrence signals (sweetness, saltiness) through mixed provenance (umami, protein, fat) to dimensions with no chemical representation at all (texture, culture, geography). Recipe co-occurrence, it turns out, encodes not just what goes with what, but \emph{why}.

This structure was always present in the raw FlavorGraph embeddings. Our LLM-augmented curation pipeline does not create new signal; it removes the noise that obscures existing signal. The raw space is contaminated by non-food entries, inconsistent naming (138 variants of ``beef''), and the resulting population imbalance in which heavily-duplicated categories dominate every aggregate statistic. Consolidation into 1,032 canonical entries produces a cleaner measurement context, improving correlations on the majority of dimensions tested, but the embeddings themselves are unchanged. This distinction matters: it implies that the structure we measure is a lower bound, and that retraining on a clean vocabulary would likely yield substantially stronger results.

The practical consequence is that ingredient embeddings could operate on interpretable axes rather than opaque similarity scores: multi-dimensional substitution, texture-aware recipe balancing, quantitative cuisine fingerprints, and clean-label reformulation all become tractable (Section~\ref{sec:discussion}). More broadly, these results suggest that the tacit knowledge embedded in culinary practice, the intuitions about flavor balance, textural contrast, and cultural coherence that chefs apply but cannot fully articulate, has computationally recoverable structure. Making that structure explicit is a step toward an in silico model of flavor that treats food not as chemistry alone, but as the integrated cultural, perceptual, and chemical system it actually is.

% ============================================================
\section*{Availability}

The \textsc{Epicure} system is available at \url{https://epicure.kaikaku.ai}. The interactive embedding explorer, which allows browsing the 1,032 curated ingredients and their similarity relationships, is available at \url{https://epicure-data.kaikaku.ai}. The curated embeddings and curation pipeline are proprietary; the LLM classification prompts, statistical methodology, and compound provenance analysis are documented in the appendix to support methodological reproducibility.

% ============================================================
\section*{Declaration of Generative AI Use}

Generative AI tools were used in the following capacities during this work:

\begin{enumerate}[leftmargin=*]
\item \textbf{Ingredient classification (core methodology).} Google Gemini~3.1~Pro was used to classify all 1,032 curated and 6,653 raw ingredients across taste, texture, binary, cultural, and ground-truth dimensions. These LLM-generated labels are the subject of the study and are described in detail in Section~\ref{sec:methods} and Appendix~\ref{sec:prompts}. All prompts, schemas, and parameters are fully documented.
\item \textbf{Data curation pipeline.} Google Gemini~2.5~Flash was used within the ingredient curation pipeline to assist with ingredient deduplication, synonym resolution, and category assignment during the reduction from 6,653 raw nodes to the 1,032 curated vocabulary (Section~\ref{sec:pipeline}).
\item \textbf{USDA/FooDB ingredient name matching (validation pipeline).} Google Gemini~2.5~Flash was used to validate automated ingredient-to-database matches (Section~\ref{sec:chemical_validation}, Appendix~\ref{sec:matching_pipeline}).
\item \textbf{Writing and code assistance.} Anthropic Claude was used as a writing aid for drafting, editing, and refactoring portions of the manuscript text and analysis code. All outputs were reviewed and verified by the authors.
\end{enumerate}

\noindent No generative AI tool was used to generate or fabricate experimental results, statistical analyses, or scientific conclusions.

% ============================================================
\bibliographystyle{plainnat}
\bibliography{references}

% ============================================================
\appendix
\section*{Appendices}
\section{Embedding Space Overview: Toroidal Manifold}
\label{sec:umap_toroid}

The 3D UMAP projection suggests that the embedding space forms a \emph{toroidal manifold}: two dense clouds, corresponding to the sweet and savoury poles, connected by a transition zone of ingredients that straddle both culinary contexts. We strongly encourage readers to explore this structure in the interactive 3D viewer at \url{https://epicure-data.kaikaku.ai}, where the spatial relationships described below can be rotated, zoomed, and inspected ingredient-by-ingredient. \emph{Caveat}: UMAP can create artificial topological features (loops, bridges) that do not exist in the original high-dimensional space~\cite{mcinnes2018umap}; the toroidal structure should be interpreted as suggestive rather than definitive until verified under alternative dimensionality reduction methods or hyperparameters. We define the sweet$\leftrightarrow$savoury axis using the LLM-assigned flavor profiles: the sweet pole centroid is the mean 3D position of ingredients classified as sweet ($n = 208$), and the savoury pole centroid is the mean position of ingredients classified as savoury ($n = 589$). Each ingredient is projected onto this axis (the \emph{along} coordinate) and assigned an angular position on the perpendicular cross-section ($\theta$).

\begin{figure}[H]
\centering
\includegraphics[width=\textwidth,page=1]{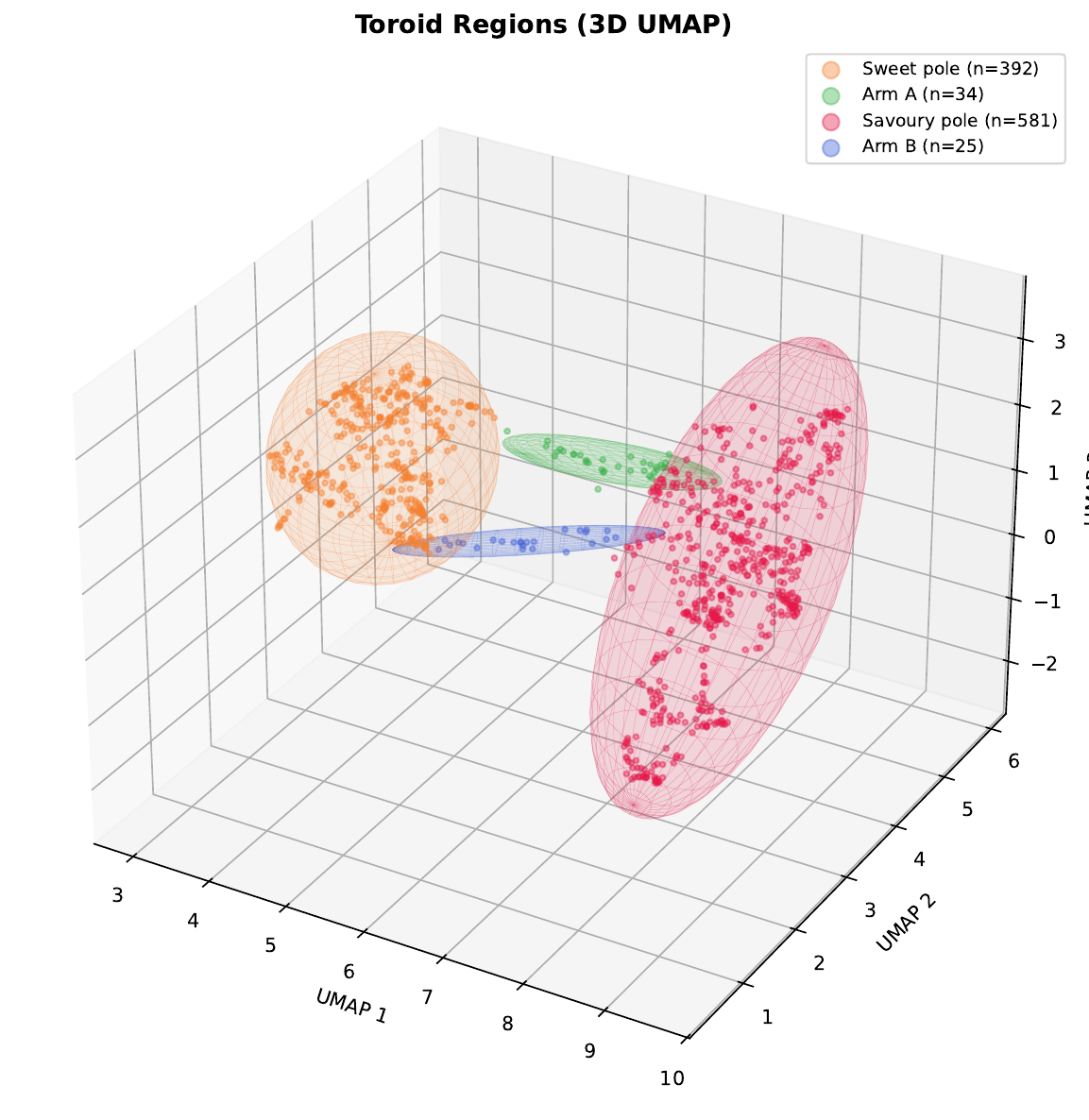}
\caption{Toroid regions in the 3D UMAP projection, with $2.5\sigma$ covariance ellipsoids showing the spatial extent of each region. The two poles (sweet, $n = 392$; savoury, $n = 581$) form dense clusters. The sweet pole is dominated by beverages, sweets, and fruits; the savoury pole by vegetables, meats, and spices.}
\label{fig:toroid_composition}
\end{figure}

Figure~\ref{fig:toroid_composition} shows the category composition of each pole. The sweet pole ($n = 392$) is dominated by Beverages (30\%), Sweets (15\%), and Fruits (14\%), with 50\% of its ingredients classified as sweet. The savoury pole ($n = 581$) is dominated by Vegetables (20\%), Condiments, Meats, and Spices ($\sim$10\% each), with 88\% classified as savoury. The transition zone between the poles contains ingredients that co-occur in recipes with both sweet and savoury dishes, reflecting their culinary versatility.

\begin{landscape}
\thispagestyle{empty}
\begin{figure}[H]
\centering
\vspace*{-0.3in}
\includegraphics[width=\linewidth,height=0.9\textheight,keepaspectratio]{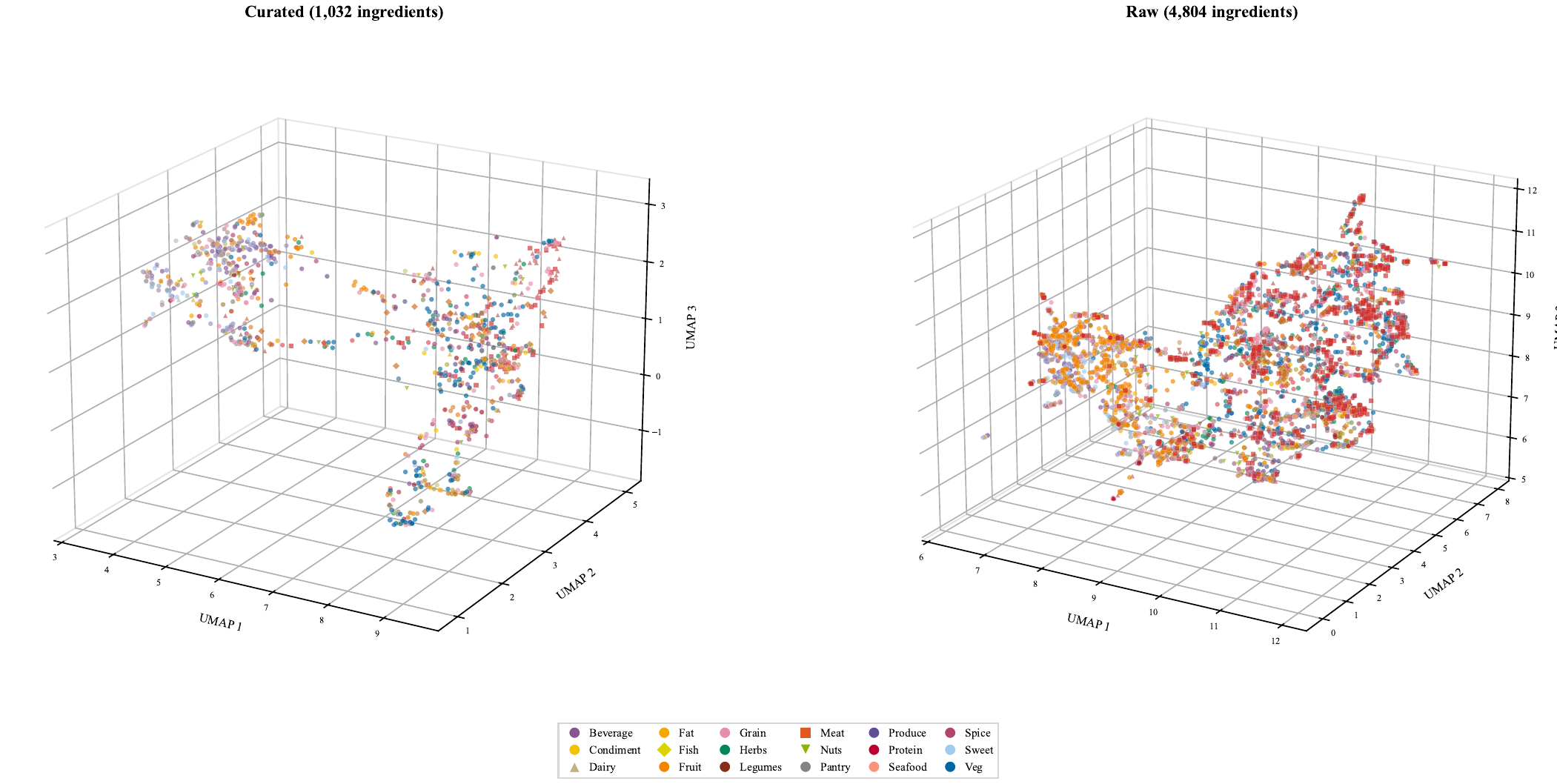}
\caption{UMAP projection from 300 dimensions to 3 dimensions. Left: curated space (1,032 ingredients). Right: raw space (6,653 ingredients, with back-projected categories). Each point is colored by the RGB average of its assigned category colors; marker shape varies by primary category. The curated space shows tighter, more distinct category clusters and a clearer toroidal structure. UMAP parameters: \texttt{n\_neighbors}$=32$, \texttt{min\_dist}$=0.004$, \texttt{metric}$=$\texttt{cosine}.}
\label{fig:umap_full}
\end{figure}
\end{landscape}

\FloatBarrier
% ============================================================
\section{LLM Classification Prompts}
\label{sec:prompts}

All ingredient classifications use Google Gemini~3.1~Pro (\texttt{gemini-3.1-pro-preview}) with structured JSON output enforced via \texttt{response\_mime\_type="application/json"} and a JSON response schema. Shared parameters: temperature~$= 0.1$, \texttt{max\_output\_tokens}~$= 16{,}000$, batch size~$= 25$ ingredients per request. Both curated (1,032) and raw (6,653) ingredient sets are tagged independently using identical prompts.

\subsection{Ground-Truth Dimensions}
\label{sec:prompt_gt}

Five dimensions: botanical family (APG~IV), climate zone, NOVA processing level, umami level, Scoville~SHU.

\begin{lstlisting}
You are a food scientist and botanist. For each ingredient
below, provide classifications across multiple dimensions. Be
accurate and conservative -- only assign a value when you are
confident.

DIMENSIONS:

1. **botanical_family**: The APG IV botanical family of the
   primary plant source.
   - Only assign for whole or minimally processed plant-derived
     ingredients where the source plant is unambiguous.
   - Use "N/A" for: animal products, highly processed items,
     multi-ingredient items, and items where the botanical
     source is ambiguous.
   - Valid families: Amaryllidaceae, Apiaceae, Asteraceae,
     Brassicaceae, Cucurbitaceae, Fabaceae, Gramineae,
     Lamiaceae, Lauraceae, Musaceae, Myrtaceae, Piperaceae,
     Poaceae, Rosaceae, Rutaceae, Solanaceae, Zingiberaceae,
     Arecaceae, Pedaliaceae, Amaranthaceae, Other, N/A

2. **climate_zone**: The primary climate zone where this
   ingredient is traditionally cultivated or originates.
   - Use "N/A" for: highly processed items, synthetic
     ingredients, items cultivated globally with no clear
     primary zone.
   - Valid zones: Tropical, Subtropical, Mediterranean,
     Temperate, Continental, Arid, Subarctic, N/A

3. **nova_level**: NOVA food processing classification
   (Monteiro et al. 2019).
   - "1" = Unprocessed or minimally processed
   - "2" = Processed culinary ingredients
   - "3" = Processed foods
   - "4" = Ultra-processed

4. **umami_level**: Free glutamate content / umami intensity.
   - "none" / "low" / "moderate" / "high" / "very_high"
   [with per-level examples as in Section 2.2]

5. **scoville_shu**: Estimated median Scoville Heat Units.
   - 0 for any ingredient that is not a significant source of
     pungent heat.

INGREDIENTS TO CLASSIFY:
[numbered list of 25 ingredients]

Return a JSON array with one object per ingredient, in the
same order as listed above.
\end{lstlisting}

\textbf{Response schema} (enforced via Gemini structured output):
\begin{lstlisting}
{ "type": "ARRAY", "items": { "type": "OBJECT",
  "properties": {
    "botanical_family": {"type":"STRING", "enum":[22 values]},
    "climate_zone":     {"type":"STRING", "enum":[8 values]},
    "nova_level":       {"type":"STRING", "enum":["1".."4"]},
    "umami_level":      {"type":"STRING",
                         "enum":["none".."very_high"]},
    "scoville_shu":     {"type":"INTEGER"}
  },
  "required": ["botanical_family","climate_zone",
               "nova_level","umami_level","scoville_shu"]
}}
\end{lstlisting}

\subsection{Taste Dimensions}
\label{sec:prompt_taste}

Four additional taste dimensions (sweet, salty, sour, bitter) appended to the ground-truth tags.

\begin{lstlisting}
You are a food scientist specializing in taste perception. For
each ingredient below, classify its intensity along 4 basic
taste dimensions. Rate the inherent taste of the ingredient
itself (not how it is used in recipes).

DIMENSIONS:

1. **sweet_level**: Perceived sweetness intensity.
   - "none" = no sweetness (salt, vinegar, most raw meats)
   - "low" = faint sweetness (milk, carrot, corn)
   - "moderate" = clearly sweet (apple, beet, sweet potato)
   - "high" = distinctly sweet (honey, maple syrup, banana)
   - "very_high" = intensely sweet (sugar, molasses, candy)

2. **salty_level**: Perceived saltiness intensity.
   [analogous 5-level scale with examples]

3. **sour_level**: Perceived sourness / acidity.
   [analogous 5-level scale with examples]

4. **bitter_level**: Perceived bitterness intensity.
   [analogous 5-level scale with examples]
\end{lstlisting}

\subsection{Texture Dimensions}
\label{sec:prompt_texture}

Six texture dimensions grounded in ISO~11036~\cite{iso11036} and the Szczesniak classification~\cite{szczesniak1963texture}. Ingredients are classified \emph{as typically used in cooking}.

\begin{lstlisting}
You are a food scientist specializing in sensory texture
profiling (ISO 11036 / Szczesniak framework). For each
ingredient below, classify its texture properties **as the
ingredient is typically used in cooking**.

TEXTURE DIMENSIONS:

1. **hardness**: Force required to compress.
   "liquid" / "gel" / "soft" / "firm" / "hard" / "very_hard"

2. **viscosity**: Resistance to flow (liquids only).
   "thin" / "slightly_thick" / "thick" / "very_thick" / "N/A"

3. **crunchiness**: Audible fracture / crisp bite.
   "none" / "slight" / "moderate" / "high" / "very_high"

4. **chewiness**: Duration of mastication before swallowing.
   "none" / "low" / "moderate" / "high" / "very_high"

5. **moisture**: Perceived water content.
   "dry" / "slightly_moist" / "moist" / "wet" / "very_wet"

6. **fattiness**: Perceived fat/oil content.
   "none" / "low" / "moderate" / "high" / "very_high"
\end{lstlisting}

Each level includes concrete food examples (e.g., hardness: ``liquid'' = water, oil, soy sauce; ``very\_hard'' = whole nutmeg, cinnamon stick, rock sugar).

\subsection{Binary Classification}
\label{sec:prompt_binary}

Independent yes/no classification for 7 dimensions, with no reference to the ordinal scales above. Uses the same model and temperature for fair comparison.

\begin{lstlisting}
You are a food scientist. For each ingredient below, answer
YES or NO for each question. Classify the ingredient **as it
is typically used in cooking**.

QUESTIONS:

1. **sour**: Is this ingredient notably sour or acidic?
   (lemon, vinegar = yes; sugar, butter = no)
2. **bitter**: Is this ingredient notably bitter?
   (coffee, dark chocolate = yes; sugar, milk = no)
3. **hard**: Is this ingredient hard or very hard?
   (raw nuts, hard candy = yes; butter, yogurt = no)
4. **crunchy**: Is this ingredient notably crunchy or crispy?
   (raw carrot, tortilla chip = yes; milk, cheese = no)
5. **chewy**: Is this ingredient notably chewy?
   (beef jerky, caramel = yes; water, sugar = no)
6. **moist**: Is this ingredient notably moist or wet?
   (watermelon, tomato = yes; flour, crackers = no)
7. **fatty**: Is this ingredient notably fatty or oily?
   (butter, olive oil = yes; water, vinegar = no)
\end{lstlisting}

\subsection{Cultural Cuisine Annotations}
\label{sec:prompt_cuisine}

Distinctive cultural marker tagging with the critical instruction that most ingredients should receive \emph{zero} cuisine tags.

\begin{lstlisting}
You are a culinary anthropologist and food scientist. For each
ingredient below, provide two classifications.

## 1. Cuisine associations -- DISTINCTIVE MARKERS ONLY

Assign cuisine tags ONLY if the ingredient is a **distinctive
cultural marker**: one that uniquely signals a specific
culinary tradition. The key test is: "If I see this ingredient
in a dish, does it immediately tell me which cuisine family
the dish belongs to?"

CRITICAL: Most ingredients should have ZERO cuisine tags.
Universal ingredients that appear across many world cuisines
must be left untagged (empty list). Examples of universal/
untagged: garlic, chicken, butter, rice, onion, salt, pepper,
olive oil, tomato, egg, sugar, flour, milk.

The 7 macro-regional cuisine clusters:
- Japanese: (dashi, miso, wasabi, nori, bonito, yuzu, mochi)
- East_Asian: Chinese + Korean (gochujang, hoisin_sauce, ...)
- Southeast_Asian: Thai + Vietnamese + Filipino + Indonesian
  (lemongrass, galangal, fish_sauce, pandan, kaffir_lime, ...)
- South_Asian: Indian + Pakistani + Sri Lankan + Bangladeshi
  (garam_masala, ghee, paneer, curry_leaf, fenugreek, ...)
- Latin_American: Mexican + Brazilian + Peruvian + Caribbean
  (chipotle, tomatillo, epazote, achiote, plantain, ...)
- Mediterranean: Italian + French + Iberian + Greek + Levant
  + North African (za'atar, harissa, prosciutto, feta, ...)
- Northern_Atlantic: Scandinavian + British + German + E.
  European + American (ranch, BBQ_sauce, horseradish, rye, ...)

## 2. Flavor profile
- "sweet" / "savoury" / "both" / "neutral"
\end{lstlisting}

% ============================================================
\section{Ingredient Matching Pipeline}
\label{sec:matching_pipeline}

Section~\ref{sec:chemical_methods} summarizes the three-layer pipeline that matches the 1,032 curated ingredients to USDA FoodData Central and FooDB entries. We provide additional detail here.

\textbf{Layer~1: Rule-based matching.} An inverted index is built over normalized segments of all database descriptions (e.g., ``Spices, cumin seed'' generates keys \texttt{spices}, \texttt{cumin seed}, \texttt{cumin}, and their stemmed forms). For each ingredient, all matching database entries are retrieved, then scored by: (a)~exact match ($\text{score} = 1000$), (b)~match after removing processing words ($900$), (c)~stemmed match ($800$), (d)~match via consolidation map ($700$), (e)~substring containment ($600-$), (f)~word overlap ($500 \times \text{Jaccard}$). Among tied candidates, a preparation-state preference order breaks ties: raw~$>$~fresh~$>$~dried~$>$~cooked~$>$~canned~$>$~frozen (with intermediate states ground, whole, boiled, roasted, baked interpolated). Approximately 60 synonym mappings handle British/American variants (e.g., courgette~$\to$~zucchini, aubergine~$\to$~eggplant, coriander leaf~$\to$~cilantro) and USDA naming conventions (e.g., ``blue cheese''~$\to$~``cheese, blue'').

\textbf{Layer~2: Embedding similarity.} Unmatched ingredients are embedded using Gemini text embeddings (\texttt{gemini-embedding-001}, 3,072 dimensions). The top-5 database entries by cosine similarity (threshold $\geq 0.80$) are forwarded to Layer~3.

\textbf{Layer~3: LLM validation.} Candidate matches from Layer~2 are validated by Gemini~2.5~Flash with the following prompt:

\begin{lstlisting}
You are an expert food scientist. Match the ingredient
"[INGREDIENT]" to one of these database entries:

Candidates: [list of candidate descriptions]

Return the BEST matching entry exactly as provided, or
"no_match" if none represent the same food. USDA names
have format "Category, specific name, preparation"
(e.g., "Spices, cumin seed" = cumin).
\end{lstlisting}

The response schema enforces \texttt{best\_match} (string) and \texttt{reasoning} (string) fields.

\textbf{Match rates.} USDA: 712/1,032 (69.0\%). FooDB: 770/1,032 (74.6\%). Unmatched ingredients are predominantly culturally specific items (e.g., gochujang, ras el hanout), composite products (e.g., Worcestershire sauce), or items not present in the database releases used.

% ============================================================
\section{FlavorGraph Training Compound Provenance}
\label{sec:compound_provenance}

FlavorGraph~\cite{park2021flavorgraph} includes 1,561 flavor compounds (a ``food'' subset sourced from FlavorDB~\cite{garg2018flavordb}) and connects them to 416 hub ingredients via 35,440 ingredient--compound edges. We classify these compounds to trace the chemical provenance of each validated dimension.

\textbf{Amino acids (12 compounds, 189 edges, 0.5\% of compound edges):} D-Isoleucine methyl ester HCl, DL-Valine, DL-Alanine, DL-Methionine, DL-Phenyl\-alanine, L-Arginine, L-Aspartic acid, L-Glutamic acid, L-Histidine, L-Lysine, L-Phenyl\-alanine, S-allyl-L-Cysteine. These connect to 75 hub ingredients.

\textbf{Fatty acids (35 compounds, 1,289 edges, 3.6\% of compound edges):} Straight-chain saturated (butyric through stearic acid, C4--C18), monounsaturated (oleic acid), polyunsaturated (linoleic, linolenic, arachidonic acid), and branched-chain aliphatic acids (2-methylbutanoic acid through 5-methylhexanoic acid). These connect to 302 hub ingredients.

\textbf{Remaining compounds (1,514, 95.9\% of edges):} Predominantly volatile aroma molecules (esters, aldehydes, ketones, terpenes, furanones, pyrazines).

\textbf{Key absences.} Sucrose, glucose, fructose, sodium, and water are \emph{not} present among the 1,561 training compounds. This confirms that the sweet and salty embedding signals arise entirely from recipe co-occurrence, while umami, protein, and fat signals have a partial chemical pathway through amino acid and fatty acid edges.

% ============================================================
\section{Cross-Validation Procedure}
\label{sec:cv_procedure}

To quantify axis-definition overfitting, we report 10-fold cross-validated estimates with 20 random repeats (200 total measurements per dimension).

\begin{algorithm}[H]
\caption{Cross-Validated Axis Projection}
\label{alg:cv}
\begin{algorithmic}[1]
\Require Normalized embeddings $\mathbf{E} \in \mathbb{R}^{n \times 300}$, labels $\mathbf{y}$, $k = 10$, $R = 20$
\Ensure Mean and std of metric across folds
\For{$r = 1$ \textbf{to} $R$}
  \State Randomly permute indices $\{1, \ldots, n\}$
  \For{fold $f = 1$ \textbf{to} $k$}
    \State Split into train $\mathcal{T}_f$ and test $\mathcal{H}_f$
    \State \textbf{Define axis from train only:}
    \If{ordinal dimension}
      \State $\mathbf{a}_f \gets \text{centroid}(\text{high-pole in } \mathcal{T}_f) - \text{centroid}(\text{low-pole in } \mathcal{T}_f)$
    \ElsIf{binary dimension}
      \State $\mathbf{a}_f \gets \text{centroid}(\text{Yes in } \mathcal{T}_f) - \text{centroid}(\text{No in } \mathcal{T}_f)$
    \ElsIf{nutritional dimension}
      \State $\mathbf{a}_f \gets \text{centroid}(\text{top tercile in } \mathcal{T}_f) - \text{centroid}(\text{bottom tercile in } \mathcal{T}_f)$
    \EndIf
    \State $\hat{\mathbf{a}}_f \gets \mathbf{a}_f / \|\mathbf{a}_f\|$ \Comment{Unit normalize}
    \State Project test: $p_i \gets \mathbf{E}_i \cdot \hat{\mathbf{a}}_f \quad \forall i \in \mathcal{H}_f$
    \If{ordinal or nutritional}
      \State $\text{metric}_{r,f} \gets \text{Spearman}(\mathbf{y}_{\mathcal{H}_f}, \mathbf{p}_{\mathcal{H}_f})$
    \Else
      \State $\text{metric}_{r,f} \gets \text{Cohen's } d(\mathbf{p}_{\text{No}}, \mathbf{p}_{\text{Yes}})$
    \EndIf
  \EndFor
\EndFor
\State \Return $\text{mean}(\text{metric}), \text{std}(\text{metric})$
\end{algorithmic}
\end{algorithm}

For ordinal dimensions, pole categories are the extreme levels: ``none'' vs.\ ``very\_high'' for tastes, ``1'' vs.\ ``4'' for NOVA, lowest vs.\ highest tercile for Scoville (log-transformed) and latitude. The axis direction is thus defined from a subset of ingredients in each fold, while the Spearman correlation is computed over \emph{all} held-out ingredients regardless of their label. For nutritional dimensions, tercile boundaries are recomputed on each training fold from USDA laboratory measurements.

The consolidated shrinkage estimates are reported inline in Tables~\ref{tab:gt_ordinal}--\ref{tab:nutritional} throughout the main text.

% ============================================================
\section{UMAP Hyperparameter Selection}
\label{sec:umap_params}

The UMAP projection used for visualization (Appendix~\ref{sec:umap_toroid}) was selected via grid search over 16 configurations:

\begin{table}[H]
\centering
\caption{UMAP grid search configurations. The selected configuration (bold) maximizes mean intra-category cosine similarity in the 3D projection.}
\label{tab:umap_grid}
\begin{tabular}{@{}llp{5cm}@{}}
\toprule
\textbf{Parameter} & \textbf{Values tested} & \textbf{Rationale} \\
\midrule
\texttt{n\_neighbors} & 15, \textbf{32}, 50, 100 & Controls local vs.\ global structure \\
\texttt{min\_dist} & 0.001, \textbf{0.004}, 0.01, 0.05 & Controls cluster tightness \\
\texttt{metric} & cosine (fixed) & Matches embedding similarity metric \\
\texttt{n\_components} & 3 (fixed) & 3D visualization \\
\bottomrule
\end{tabular}
\end{table}

The selection criterion was \emph{tightest category separation}: the configuration maximizing the ratio of mean within-category cosine similarity to mean cross-category cosine similarity in the 3D projection space. Final parameters: \texttt{n\_neighbors}~$= 32$, \texttt{min\_dist}~$= 0.004$, \texttt{metric}~$=$~\texttt{cosine}. UMAP is used for visualization only; all quantitative claims in the paper are computed in the native 300-dimensional space.

% ============================================================
\section{Example Ingredient Classifications}
\label{sec:example_classifications}

Table~\ref{tab:spot_check} shows classifications for 20 representative ingredients, illustrating the range and quality of LLM-assigned labels across taste, texture, and cultural dimensions.

\begin{landscape}
\begin{table}[p]
\centering
\caption{Spot-check of LLM classifications across 20 representative ingredients. Sweet/Salty/Umami: ordinal levels (n/l/m/h/vh = none/low/moderate/high/very\_high). SHU: Scoville Heat Units. NOVA: processing level (1--4). Binary columns: direct yes/no LLM labels. Cuisines: distinctive cultural marker tags (Jp = Japanese, EA = East Asian, SEA = Southeast Asian, Med = Mediterranean, -- = universal). FP: flavor profile (sw = sweet, sv = savoury, b = both, n = neutral).}
\label{tab:spot_check}
\scriptsize
\begin{tabular}{@{}l ccc r c cc cc ccccc ll @{}}
\toprule
& \multicolumn{3}{c}{\textbf{Ordinal taste}} & \textbf{SHU} & \textbf{NOVA}
& \multicolumn{2}{c}{\textbf{Texture}} & \multicolumn{2}{c}{\textbf{Binary taste}}
& \multicolumn{5}{c}{\textbf{Binary texture}} & & \\
\cmidrule(lr){2-4} \cmidrule(lr){7-8} \cmidrule(lr){9-10} \cmidrule(lr){11-15}
\textbf{Ingredient} & Sw & Sa & Um & & & Hard & Moist & Sour & Bitter & Hard & Crunch & Moist & Fatty & Chewy & \textbf{Cuisines} & \textbf{FP} \\
\midrule
sugar         & vh & n  & n  & 0      & 2 & hard   & dry   & no  & no  & no & no & no & no & no & --      & sw \\
soy sauce     & n  & h  & h  & 0      & 3 & liquid & v.wet & no  & no  & no & no & yes & no & no & Jp,EA,SEA & sv \\
miso paste    & l  & h  & h  & 0      & 3 & gel    & wet   & no  & no  & no & no & yes & no & no & Jp      & sv \\
cayenne       & n  & n  & n  & 40k    & 1 & soft   & dry   & no  & no  & no & no & no  & no & no & --      & sv \\
wasabi        & l  & n  & l  & 0      & 1 & gel    & wet   & no  & no  & no & no & yes & no & no & Jp      & sv \\
lemon         & n  & n  & n  & 0      & 1 & soft   & v.wet & yes & no  & no & no & yes & no & no & --      & b  \\
coffee        & n  & n  & n  & 0      & 1 & liquid & v.wet & no  & yes & no & no & yes & no & no & --      & b  \\
dark choc.    & l  & n  & l  & 0      & 3 & hard   & dry   & no  & yes & yes & no & no  & yes & no & --     & sw \\
butter        & l  & l  & n  & 0      & 2 & soft   & sl.m  & no  & no  & no & no & no  & yes & no & --      & b  \\
olive oil     & n  & n  & n  & 0      & 2 & liquid & v.wet & no  & no  & no & no & yes & yes & no & --      & sv \\
watermelon    & m  & n  & n  & 0      & 1 & soft   & v.wet & no  & no  & no & no & yes & no & no & --      & sw \\
parmesan      & l  & h  & h  & 0      & 3 & hard   & sl.m  & no  & no  & yes & no & no  & yes & no & Med     & sv \\
fish sauce    & l  & vh & h  & 0      & 3 & liquid & v.wet & no  & no  & no & no & yes & no & no & SEA     & sv \\
gochujang     & m  & h  & h  & 2k     & 3 & liquid & moist & no  & no  & no & no & yes & no & no & EA      & sv \\
lemongrass    & n  & n  & n  & 0      & 1 & hard   & sl.m  & no  & no  & yes & no & yes & no & no & SEA     & sv \\
cumin         & n  & n  & n  & 0      & 1 & soft   & dry   & no  & yes & no & no & no  & no & no & --      & sv \\
vanilla       & n  & n  & n  & 0      & 1 & liquid & v.wet & no  & yes & no & no & yes & no & no & --      & sw \\
tofu          & l  & n  & m  & 0      & 3 & soft   & wet   & no  & no  & no & no & yes & no & no & EA,Jp,SEA & n  \\
almond        & l  & n  & n  & 0      & 1 & hard   & sl.m  & no  & no  & yes & no & no  & yes & no & --      & b  \\
\bottomrule
\end{tabular}
\end{table}
\end{landscape}

Notable classifications: (1)~Wasabi correctly receives 0~SHU, its pungency is isothiocyanate-based, not capsaicin. (2)~Parmesan is tagged high umami, high salty, hard texture, and Mediterranean cuisine, consistent with its culinary role. (3)~Universal ingredients (sugar, lemon, butter, cumin) receive no cuisine tags, while culturally distinctive items (miso, gochujang, fish sauce, lemongrass) receive specific tags. (4)~Tofu is tagged across three Asian cuisine families, reflecting its pan-Asian usage.

\end{document}